\documentclass[reprint,superscriptaddress,aps]{revtex4-1}
\usepackage{amsmath}
\usepackage{amssymb}
\usepackage{bm}
\usepackage{braket}
\usepackage{color}
\usepackage[varg]{txfonts}
\usepackage{varwidth}
\usepackage{dcolumn}
\usepackage[breaklinks,colorlinks=true,linkcolor=blue,urlcolor=cyan,citecolor=blue]{hyperref}
\usepackage{here}
\usepackage{graphicx}

\begin{document}

\title{Rhenium oxyhalides: a showcase for anisotropic-triangular-lattice quantum antiferromagnets}

\author{M. Gen}
\email{masaki.gen@riken.jp}
\affiliation{Institute for Solid State Physics, The University of Tokyo, Kashiwa, Chiba 277-8581, Japan}
\affiliation{RIKEN Center for Emergent Matter Science (CEMS), Wako 351-0198, Japan}

\author{D. Hirai}
\affiliation{Institute for Solid State Physics, The University of Tokyo, Kashiwa, Chiba 277-8581, Japan}
\affiliation{Department of Applied Physics, Nagoya University, Nagoya 464-8603, Japan}

\author{K. Morita} 
\affiliation{Department of Physics, Faculty of Science and Technology, Tokyo University of Science, Chiba 278-8510, Japan}

\author{S. Kogane}
\affiliation{Institute for Solid State Physics, The University of Tokyo, Kashiwa, Chiba 277-8581, Japan}

\author{N. Matsuyama}
\affiliation{Institute for Solid State Physics, The University of Tokyo, Kashiwa, Chiba 277-8581, Japan}

\author{T. Yajima}
\affiliation{Institute for Solid State Physics, The University of Tokyo, Kashiwa, Chiba 277-8581, Japan}

\author{M. Kawamura}
\affiliation{Institute for Solid State Physics, The University of Tokyo, Kashiwa, Chiba 277-8581, Japan}
\affiliation{Information Technology Center, The University of Tokyo, Tokyo 113-8658, Japan}

\author{K. Deguchi}
\affiliation{Department of Physics, Nagoya University, Nagoya 464-8602, Japan}

\author{A. Matsuo}
\affiliation{Institute for Solid State Physics, The University of Tokyo, Kashiwa, Chiba 277-8581, Japan}

\author{K. Kindo}
\affiliation{Institute for Solid State Physics, The University of Tokyo, Kashiwa, Chiba 277-8581, Japan}

\author{Y. Kohama}
\affiliation{Institute for Solid State Physics, The University of Tokyo, Kashiwa, Chiba 277-8581, Japan}

\author{Z. Hiroi}
\affiliation{Institute for Solid State Physics, The University of Tokyo, Kashiwa, Chiba 277-8581, Japan}

\begin{abstract}

The spin-1/2 Heisenberg antiferromagnet on an anisotropic triangular lattice (ATL) is an archetypal spin system hosting exotic quantum magnetism and dimensional crossover.
However, the progress in experimental research on this field has been limited due to the scarcity of ideal model materials.
Here, we show that rhenium oxyhalides {\it A}$_{3}$ReO$_{5}${\it X}$_{2}$, where spin-1/2 Re$^{6+}$ ions form a layered structure of ATLs, allow for flexible chemical substitution in both cation {\it A}$^{2+}$ ({\it A} = Ca, Sr, Ba, Pb) and anion {\it X}$^{-}$ ({\it X} = Cl, Br) sites, leading to seven synthesizable compounds.
By combining magnetic susceptibility and high-field magnetization measurements with theoretical calculations using the orthogonalized finite-temperature Lanczos method, we find that the anisotropy $J'/J$ ranges from 0.25 to 0.45 depending on the chemical composition.
Our findings demonstrate that {\it A}$_{3}$ReO$_{5}${\it X}$_{2}$ is an excellent platform for realizing diverse effective spin Hamiltonians that differ in the strength of the anisotropy $J'/J$ as well as the relevance of perturbation terms such as the Dzyaloshinskii-Moriya interaction and interlayer exchange coupling.

\end{abstract}

\date{\today}
\maketitle

Frustrated quantum magnets have long drawn considerable attention due to the possibility of realizing exotic magnetic states \cite{2010_Bal} and fractional quasi-particle excitations \cite{2007_Koh}.
Although recent advances in analytical and numerical approaches to describe many-body quantum physics have promoted our understanding of various magnetic phenomena, many issues remain unresolved \cite{2016_Nor}, highlighting the need for further experimental insights.
To this end, it is desirable to develop model compounds with a simple effective spin Hamiltonian, an accessible energy scale (i.e., temperature and magnetic field), and a broad tunability of perturbative terms as well as exchange couplings through chemical substitution.

The spin-1/2 Heisenberg antiferromagnet on an anisotropic triangular lattice (ATL) is an archetypal frustrated spin system \cite{1999_Wei, 2003_Chu, 2004_Yun, 2005_Zhe, 2006_Zhe_1, 2006_Yun, 2006_Wen, 2007_Sta, 2007_Hay, 2009_Hei, 2010_Sta, 2011_Reu, 2012_Har, 2014_Sta, 2016_Gho, 2022_Mor, 2023_Yu}.
The minimal model consists of two kinds of antiferromagnetic (AFM) exchange couplings, $J$ and $J'$, which dominate between the nearest-neighbor sites along the intrachain and interchain directions, respectively.
By varying the ratio $J'/J$, this model provides an interpolation between a regular triangular lattice ($J'/J=1$) and a decoupled one-dimensional (1D) chain ($J'/J=0$), with known ground states of the 120$^{\circ}$ long-range order (LRO) \cite{1994_Ber, 2007_Whi} and the Tomonaga-Luttinger liquid (TLL) \cite{1974_End, 1993_Ten}, respectively.
Theoretically, it is well established that the TLL-like gapless quantum spin-liquid (QSL) is robust for a wide parameter range of $0 < J'/J \lesssim 0.6$ \cite{2006_Yun, 2009_Hei, 2016_Gho}.
This is a manifestation of the so-called dimensional reduction caused by geometrical frustration \cite{2007_Hay}.
On the other hand, the ground state for $0.6 \lesssim J'/J \lesssim 0.8$ is controversial; some theories suggest a direct transition from the TLL-like QSL to the spiral LRO \cite{2006_Wen, 2007_Sta, 2011_Reu, 2012_Har, 2016_Gho}, while others suggest the appearance of another phase, such as a dimer-ordered state \cite{1999_Wei} and a gapped QSL \cite{2006_Yun, 2009_Hei}.

Experimentally, two types of compounds have been intensively studied as candidates for the ATL quantum antiferromagnet: Cs$_{2}$Cu{\it X}$_{4}$ ({\it X} = Cl, Br) \cite{1996_Col, 2001_Col, 2002_Col, 2003_Col, 2005_Rad, 2006_Tok, 2011_Pov, 2012_Smi, 2014_Zvy, 2019_Zvy, 2019_Sch, 2003_Ono, 2004_Ono, 2007_Tsu, 2009_For} and organic Mott insulators represented by $\kappa$-(BEDT-TTF)$_{2}$Cu$_{2}$(CN)$_{3}$ \cite{2003_Shi, 2009_Kan, 2014_Kor, 2015_Yos, 1996_Kom, 1998_Kat, 2021_Shi, 2017_Pow}.
The magnetic properties of Cs$_{2}$Cu{\it X}$_{4}$ have been unveiled in detail thanks to relatively weak AFM exchange couplings: $J/k_{\rm B} = 4.3$~K and $J'/J = 0.30$ ($J/k_{\rm B} = 14.9$~K and $J'/J = 0.41$) for {\it X} = Cl (Br) \cite{2014_Zvy}, where $k_{\rm B}$ is the Boltzmann constant.
However, Cs$_{2}$Cu{\it X}$_{4}$ undergoes a magnetic transition into an incommensurate spiral LRO at low temperatures \cite{1996_Col, 2003_Ono} due to the presence of the Dzyaloshinskii-Moriya (DM) interaction \cite{2010_Sta}, which interfere with the observation of essential properties expected for the ideal ATL quantum antiferromagnet.
For the organic Mott insulators, the exchange couplings are much stronger, on the order of several hundred kelvin \cite{2005_Zhe, 2003_Shi, 2009_Kan, 2014_Kor, 2015_Yos}, making the accurate estimation of $J'/J$ elusive.
The contribution of multi-spin exchange and spin-lattice coupling can be important in the organic salts \cite{2005_Zhe}, as suggested from a pressure-induced insulator-metal transition \cite{1996_Kom, 1998_Kat} and a magnetic-field-induced AFM transition reminiscent of a spin-Peierls phase \cite{2021_Shi}, respectively.

\begin{figure*}[t]
\centering
\includegraphics[width=\linewidth]{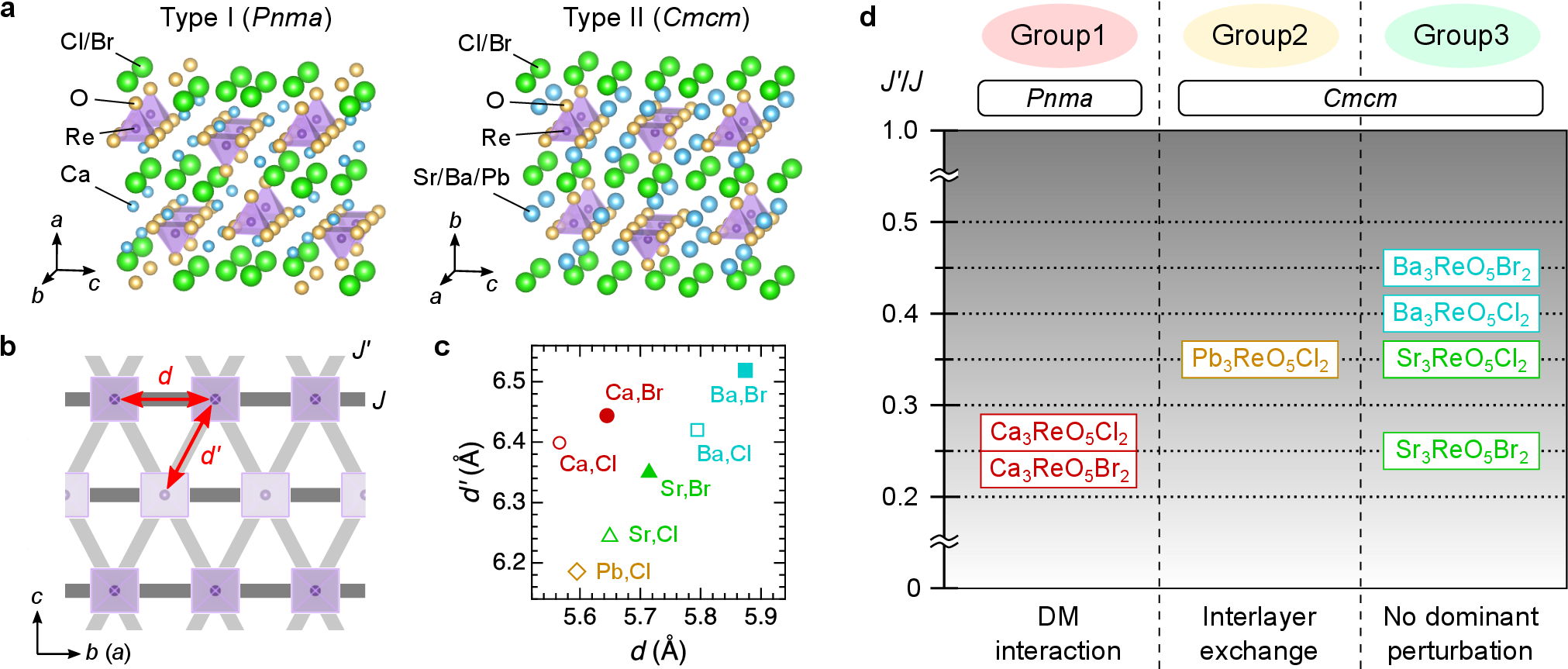}
\caption{{\bf Structural and magnetic properties of {\mbox {\boldmath $A$}}$_{3}$ReO$_{5}${\mbox {\boldmath $X$}}$_{2}$.} {\bf a} Two kinds of crystal structures. Ca$_{3}$ReO$_{5}${\it X}$_{2}$ crystallizes in Type I with the space group {\it Pnma} (left), whereas the system with {\it A} = Sr, Ba, and Pb crystallizes in Type II with the space group {\it Cmcm} (right). The illustrations are drawn with VESTA software \cite{2011_Mom}. {\bf b} Schematic of an anisotropic triangular lattice (ATL), where upward and downward ReO$_{5}$ square pyramids running along the {\it b} ({\it a}) direction are alternately arranged along the {\it c} direction in the Type I (II) structure. Exchange couplings between nearest-neighbor and next-nearest-neighbor Re ions with the bond length $d$ and $d'$ are defined as $J$ and $J'$, respectively. {\bf c} Plots of $d$ and $d'$ for the seven compounds. Elements of {\it A} and {\it X} are written next to each marker. The data of {\it A}$_{3}$ReO$_{5}$Cl$_{2}$ ({\it A} = Ca, Sr, Ba) are taken from Refs.~\cite{2017_Hir, 2020_Hir}. {\bf d} Summary of effective spin Hamiltonians with the ATL quantum Heisenberg model. The anisotropy $J'/J$ for each compound is revealed in the present work and displayed in the left axis. The DM interaction and interlayer exchange coupling dominate the low-temperature magnetism for {\it A} = Ca and Pb, respectively, whereas there are no significant effects of such perturbations for {\it A} = Sr and Ba.}
\label{Fig1}
\end{figure*}

Compared to the aforementioned compounds, recently-discovered rhenium oxychlorides {\it A}$_{3}$ReO$_{5}$Cl$_{2}$ ({\it A} = Ca, Sr, Ba) \cite{2017_Hir, 2019_Hir, 2020_Hir, 2020_Naw, 2022_Zvy, 2021_Cho} potentially hold an advantage in the search for low-temperature magnetism of the ATL quantum antiferromagnet.
The Re$^{6+}$ ion has a $5d^{1}$ electronic configuration, and the unique crystal field resulting from the surrounding mixed anions causes the $d_{xy}$ orbital to become the lowest-energy level without degeneracy.
As a consequence, {\it A}$_{3}$ReO$_{5}$Cl$_{2}$ can be regarded as a spin-1/2 Heisenberg system with quenched orbital angular momentum.
$J'/J$ is estimated to be 0.32, 0.43, and 0.47 for {\it A} = Ca, Sr, and Ba, respectively, based on theoretical fits to the temperature dependence of magnetic susceptibility using high-temperature series expansion \cite{2019_Hir, 2020_Hir}.
{\it A}$_{3}$ReO$_{5}$Cl$_{2}$ exhibits TLL-like characteristics indicative of one-dimensionality at low temperatures, as evidenced by several experimental observations such as the Bonner--Fisher-like magnetic susceptibility \cite{1964_Bon}, large $T$-linear magnetic heat capacity \cite{2019_Hir, 2020_Hir}, and spinon and triplon excitations in inelastic neutron scattering \cite{2020_Naw} and Raman spectra \cite{2021_Cho}.
The frustration factor, defined as $f \equiv |\Theta_{\rm W}|/T_{\rm N}$, where $\Theta_{\rm W}$ and $T_{\rm N}$ represent a Weiss temperature and an ordering temperature, respectively, is larger in {\it A}$_{3}$ReO$_{5}$Cl$_{2}$ than in Cs$_{2}$Cu{\it X}$_{4}$ \cite{2019_Hir, 2020_Hir}.
This suggests that additional interactions, aside from $J$ and $J'$, play a less significant role in {\it A}$_{3}$ReO$_{5}$Cl$_{2}$ than in Cs$_{2}$Cu{\it X}$_{4}$.
Further development of chemical substitutes for {\it A}$_{3}$ReO$_{5}$Cl$_{2}$, along with the evaluation of their effective spin models, should open up the route for a systematic understanding of the fundamental properties of the ATL quantum antiferromagnet.

Here, we report physical properties of four novel Re-based quantum ATL antiferromagnets, {\it A}$_{3}$ReO$_{5}$Br$_{2}$ ({\it A} = Ca, Sr, Ba) and Pb$_{3}$ReO$_{5}$Cl$_{2}$.
These compounds are (nearly) isostructural with {\it A}$_{3}$ReO$_{5}$Cl$_{2}$ (Fig.~\ref{Fig1}a--c) and exhibit one-dimensional character at low temperatures as well \cite{2017_Hir, 2019_Hir, 2020_Hir}.
We also investigate magnetization processes of all the seven {\it A}$_{3}$ReO$_{5}${\it X}$_{2}$ families in pulsed high magnetic fields of up to 130~T \cite{2020_Gen}.
On the basis of theoretical calculations using the orthogonalized finite-temperature Lanczos method (OFTLM) \cite{2020_Mor, 2022_Mor}, $J'/J$ is found to vary between 0.25 and 0.45 from compound to compound.
We propose that, considering the difference in their crystal structures and spatial distributions of electronic density, {\it A}$_{3}$ReO$_{5}${\it X}$_{2}$ can be categorized into three groups in terms of an effective spin Hamiltonian (Fig.~\ref{Fig1}d): (1) {\it A} = Ca with the DM interaction, (2) {\it A} = Pb with the relatively strong interlayer exchange coupling, and (3) {\it A} = Sr and Ba with no dominant perturbations.

\section*{Results}

\vspace{-0.2cm}
\subsection*{Crystal structures}
\vspace{-0.2cm}

\begin{figure*}[t]
\centering
\includegraphics[width=\linewidth]{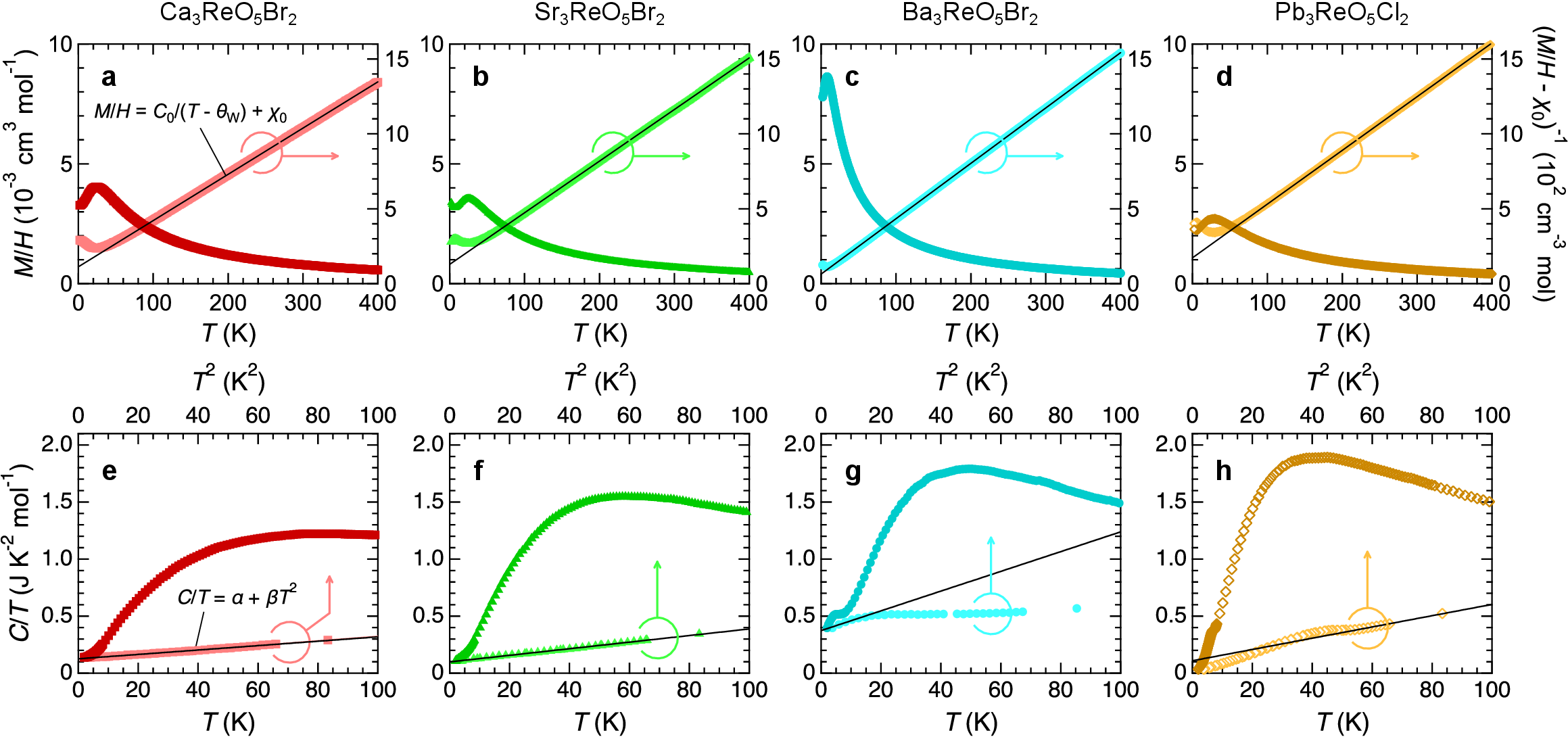}
\caption{{\bf Basic physical properties of four novel compounds.} {\bf a--d}  Temperature dependence of the magnetic susceptibility $M/H$ measured at 7~T (left axis) and the inverse susceptibility $(M/H - \chi_{0})^{-1}$ (right axis), where a $T$-independent component $\chi_{0}$ is subtracted, for CROB ({\bf a}), SROB ({\bf b}), BROB ({\bf c}), and PROC ({\bf d}). The black lines represent the Curie--Weiss law, i.e., $M/H = C/(T-\Theta_{\rm W}) + \chi_{0}$, fitted to the experimental data between 150 and 400~K. {\bf e--h} Heat capacity divided by temperature $C/T$ as a function of $T$ below 100~K (bottom axis) and as a function of $T^{2}$ below 10~K (top axis) for CROB ({\bf e}), SROB ({\bf f}), BROB ({\bf g}), and PROC ({\bf h}). The black lines represent fittings $C/T = \alpha + \beta T^{2}$ to the experimental data between 2 and 10~K for CROB and SROB, between 2 and 4~K for BROB, and between 7 and 10~K for PROC.}
\label{Fig2}
\end{figure*}

We first present the crystal structures of newly synthesized compounds: Ca$_{3}$ReO$_{5}$Br$_{2}$ (CROB), Sr$_{3}$ReO$_{5}$Br$_{2}$ (SROB), Ba$_{3}$ReO$_{5}$Br$_{2}$ (BROB), and Pb$_{3}$ReO$_{5}$Cl$_{2}$ (PROC).
Regarding the previously reported oxychlorides, Ca$_{3}$ReO$_{5}$Cl$_{2}$ (CROC) crystallizes in an orthorhombic structure with the space group $Pnma$ (No.~62) \cite{2017_Hir}, whereas Sr$_{3}$ReO$_{5}$Cl$_{2}$ (SROC) and Ba$_{3}$ReO$_{5}$Cl$_{2}$ (BROC) with the space group $Cmcm$ (No.~63) \cite{2020_Hir}, as depicted in Fig.~\ref{Fig1}a.
Hereafter, we refer to these two structures as Type I and Type II, respectively.
Both structures consist of alternating layers of {\it A}$_{3}$ReO$_{5}$ and Cl slabs.
The coordination numbers of the {\it A}$^{2+}$ ions are 5 or 6 in the Type I structure, whereas they are 8 or 9 in the Type II structure.
This difference arises due to the smaller ionic radius of Ca$^{2+}$ (1.00~$\AA$ in the 6-fold coordination) compared to Sr$^{2+}$ and Ba$^{2+}$ (1.26 and 1.42~$\AA$, respectively, in the 8-fold coordination).
We find that the substitution of Cl$^{-}$ (ionic radius 1.81~\AA) with Br$^{-}$ (ionic radius 1.96~\AA) does not impact the original crystal structure; CROB crystallizes in Type-I, while SROB and BROB crystallize in Type II.
PROC also crystallizes in Type-II (the ionic radius of Pb$^{2+}$ is 1.29~$\AA$ in the 8-fold coordination).
Further details on the crystal structures are provided in Supplementary Note~2.

An ATL composed of magnetic Re$^{6+}$ ions is depicted in Fig.~\ref{Fig1}b.
The ReO$_{5}$ square pyramids, which point upward and downward alternately along the interchain direction, do not share any oxygen ions.
Consequently, each Re$^{6+}$ ion is well separated from one another by more than 5~$\AA$.
Figure~\ref{Fig1}c compares the lengths of the nearest-neighbor and next-nearest-neighbor Re--Re bonds, $d$ and $d'$, respectively, among all the seven {\it A}$_{3}$ReO$_{5}${\it X}$_{2}$ compounds.
In common with {\it A} = Ca, Sr, and Ba, the substitution of Cl with Br leads to the elongation of both $d$ and $d'$; the increase rates are 1.4\% and 0.7\% for {\it A} = Ca, 1.2\% and 1.7\% for {\it A} = Sr, and 1.4\% and 1.5\% for {\it A} = Ba, respectively.

\vspace{-0.2cm}
\subsection*{One dimensionality of the low-temperature magnetism}
\vspace{-0.2cm}

Next, we show the basic physical properties of {\it A}$_{3}$ReO$_{5}$Br$_{2}$ ({\it A} = Ca, Sr, Ba) and PROC.
Figure~\ref{Fig2}a--d shows the temperature dependence of magnetization divided by a magnetic field $M/H$ measured at 7~T.
All the $M/H$ data above 150~K are well fitted by the Curie--Weiss law $M/H = C_{0}/(T-\Theta_{\rm W}) + \chi_{0}$, where $C_{0}$ is the Curie constant, and $\chi_{0}$ is the temperature-independent term, as shown by black lines in Fig.~\ref{Fig2}a--d.
The fitting parameters as well as the effective magnetic moment $\mu_{\rm eff}$ calculated from $C_{0}$ are listed in Table~\ref{Tab1}.
The obtained $\chi_{0}$ is comparable to the diamagnetic contribution of core electrons $\chi_{\rm dia}$ for each compound: $\chi_{\rm dia} = -1.76 \times 10^{-4}$, $-2.02 \times 10^{-4}$, $-2.24 \times 10^{-4}$ and $-2.18 \times 10^{-4}$~cm$^{3}$~mol$^{-1}$ for CROB, SROB, BROB, and PROC, respectively.
The reduction of $\mu_{\rm eff}$ from 1.73~$\mu_{\rm B}$ expected for the pure $S=1/2$ case would be attributed to the effect of the spin-orbit interaction, which cancels the spin and orbital angular momentum with each other in the $5d^{1}$ electronic configuration.

The large negative Weiss temperatures indicate the dominance of AFM exchange interactions in these compounds.
The smaller value of $|\Theta_{\rm W}|$ for BROB ($\Theta_{\rm W} = -17.2(2)$~K) compared to CROB ($-36.5(2)$~K) and SROB ($-37.4(4)$~K) can be attributed to the relatively long Re--Re bond lengths, $d$ and $d'$, in BROB (Fig.~\ref{Fig1}c).
This trend is common with the {\it A}$_{3}$ReO$_{5}$Cl$_{2}$ case: $\Theta_{\rm W} = -37.8(1), -49.5(1)$ and $-21.6(1)$~K for CROC, SROC, and BROC, respectively \cite{2019_Hir, 2020_Hir}.
Furthermore, the substitution of Cl with Br leads to a decrease in $|\Theta_{\rm W}|$; the decrease rates are 3.4~\%, 24.4~\%, and 20.4~\% for {\it A} = Ca, Sr, and Ba, respectively.
The much smaller decrease rate for {\it A} = Ca compared to {\it A} = Sr and Ba would be due to the difference in their crystal structures, as mentioned above (Fig.~\ref{Fig1}a).

\begin{table}[t]
\renewcommand{\arraystretch}{1.2}
\caption{Magnetic and thermal parameters for {\it A}$_{3}$ReO$_{5}${\it X}$_{2}$ ({\it A} = Ca, Sr, Ba, Pb; {\it X} = Cl, Br). Data for CROC are taken from Ref.~\cite{2019_Hir}, and those for SROC and BROC are taken from Ref.~\cite{2020_Hir}. See text for details of the definition of each parameter.}
\begin{tabular}{cccccccc} \hline\hline
~~ & CROC & SROC & BROC & PROC \\ \hline
$\mu_{\rm eff}$~($\mu_{\rm B}$) & 1.538(2) & 1.648(1) & 1.480(3) & 1.493(1) \\
$\Theta_{\rm W}$ (K) & $-37.8(1)$ & $-49.5(1)$ & $-21.6(1)$ & $-47.8(2)$ \\
~$\chi_{0}$~(10$^{-4}$~cm$^{3}$~mol$^{-1}$)~~ & $-1.361(5)$ & $-2.219(5)$ & $-1.931(7)$ & $-2.164(7)$ \\
~$\alpha$~(mJ~K$^{-2}$~mol$^{-1}$)~~ & ~114.7(4)~ & ~93.3(5)~ & ~246.7(4)~ & ~~109(2)~~ \\
~~$\beta$~(mJ~K$^{-4}$~mol$^{-1}$)~~ & 1.64(1) & 2.21(2) & 6.11(3) & 4.93(3) \\
\hline\hline
~~ & CROB & SROB & BROB & \\ \hline
$\mu_{\rm eff}$~($\mu_{\rm B}$) & 1.609(2) & 1.524(2) & 1.471(1) & \\
$\Theta_{\rm W}$ (K) & $-36.5(2)$ & $-37.4(4)$ & $-17.2(2)$ & \\
~$\chi_{0}$~(10$^{-4}$~cm$^{3}$~mol$^{-1}$)~~ & $-1.740(8)$ & $-1.936(16)$ & $-2.188(6)$ & \\
~$\alpha$~(mJ~K$^{-2}$~mol$^{-1}$)~~ & ~126.2(7)~ & ~~99.0(7)~~ & ~~373(9)~~ & \\
~$\beta$~(mJ~K$^{-4}$~mol$^{-1}$)~~ & 1.92(2) & 2.89(2) & 8.65(9) & \\
\hline\hline
\end{tabular}
\label{Tab1}
\end{table}

On further cooling below 100~K, all the $M/H$--$T$ curves gradually deviate from the Curie--Weiss behavior and exhibit a broad peak around 10--20~K (Fig.~\ref{Fig2}a--d), signaling the development of AFM short-range correlations.
In the ATL Heisenberg antiferromagnet with moderate anisotropy $J'/J$, a dimensional crossover from 2D to 1D, i.e., the so-called one-dimensionalization, is expected at low temperatures, where the interchain zigzag couplings $J'$ are effectively cancelled out due to geometrical frustration, and consequently a TLL-like disordered state would be realized \cite{2019_Hir, 2020_Hir, 2020_Naw, 2021_Cho, 2022_Zvy}.

Figure~\ref{Fig2}e--h shows the temperature dependence of heat capacity divided by temperature $C/T$ for each compound.
We also display $C/T$ versus $T^{2}$ plots in the top axes.
For {\it A}$_{3}$ReO$_{5}$Br$_{2}$, $C/T$ exhibits a linear $T^{2}$ dependence at low temperatures with a finite intercept, the origin of which would be gapless-spin excitation expected in the TLL-like disordered state.
Note that a hump structure is observed below 10~K in the $C/T$--$T$ curve for BROB (Fig.~\ref{Fig2}g).
This would reflect the entropy release caused by the development of magnetic short-range correlations rather than a magnetic LRO, as suggested from a broad peak around 8~K in the magnetic susceptibility data (Fig.~\ref{Fig2}c).
By fitting the $C/T$--$T^{2}$ curve to the equation $C/T = \alpha + \beta T^{2}$, where the first and second terms originate from magnetic and phonon contributions, respectively, in a temperature range between 2 and 10~K for CROB and SROB and between 2 and 4~K for BROB, we obtain the values $\alpha$ and $\beta$ for each compound as summarized in Table~\ref{Tab1} (Note that the estimation of $\beta$ for BROC has an uncertainty of approximately $\pm 30$~\% depending on the fitting range).
In the 1D chain limit ($J' = 0$), the $T$-linear contribution to the heat capacity is given by $2RT/3J$ \cite{1964_Bon}, where $R$ is the gas constant, so that $\alpha$ is inversely related to $J$.
Indeed, $\alpha$ in BROB is much larger than those in CROB and SROB.
This is compatible with the much smaller value of $|\Theta_{\rm W}|$ in BROB.
Furthermore, we find that $\alpha$ in {\it A}$_{3}$ReO$_{5}$Br$_{2}$ is a bit larger than that in the oxychloride counterpart {\it A}$_{3}$ReO$_{5}$Cl$_{2}$ \cite{2019_Hir, 2020_Hir} (Table~\ref{Tab1}).
This tendency also agrees with the decrease of $|\Theta_{\rm W}|$ by the substitution of Cl with Br.
In contrast, $C/T$ approaches zero toward the low-temperature limit for PROC.
This is owing to the occurrence of a magnetic LRO at $T_{\rm N} = 6.6$~K, as discussed below.
Accordingly, we perform a linear fit to the $C/T$--$T^{2}$ curve of PROC between 7 and 10~K (Table~\ref{Tab1}).
The magnitudes of ${\alpha}$ and $|\Theta_{\rm W}|$ in PROC are both intermediate between those in CROC and SROC.

\begin{figure}[t]
\centering
\includegraphics[width=\linewidth]{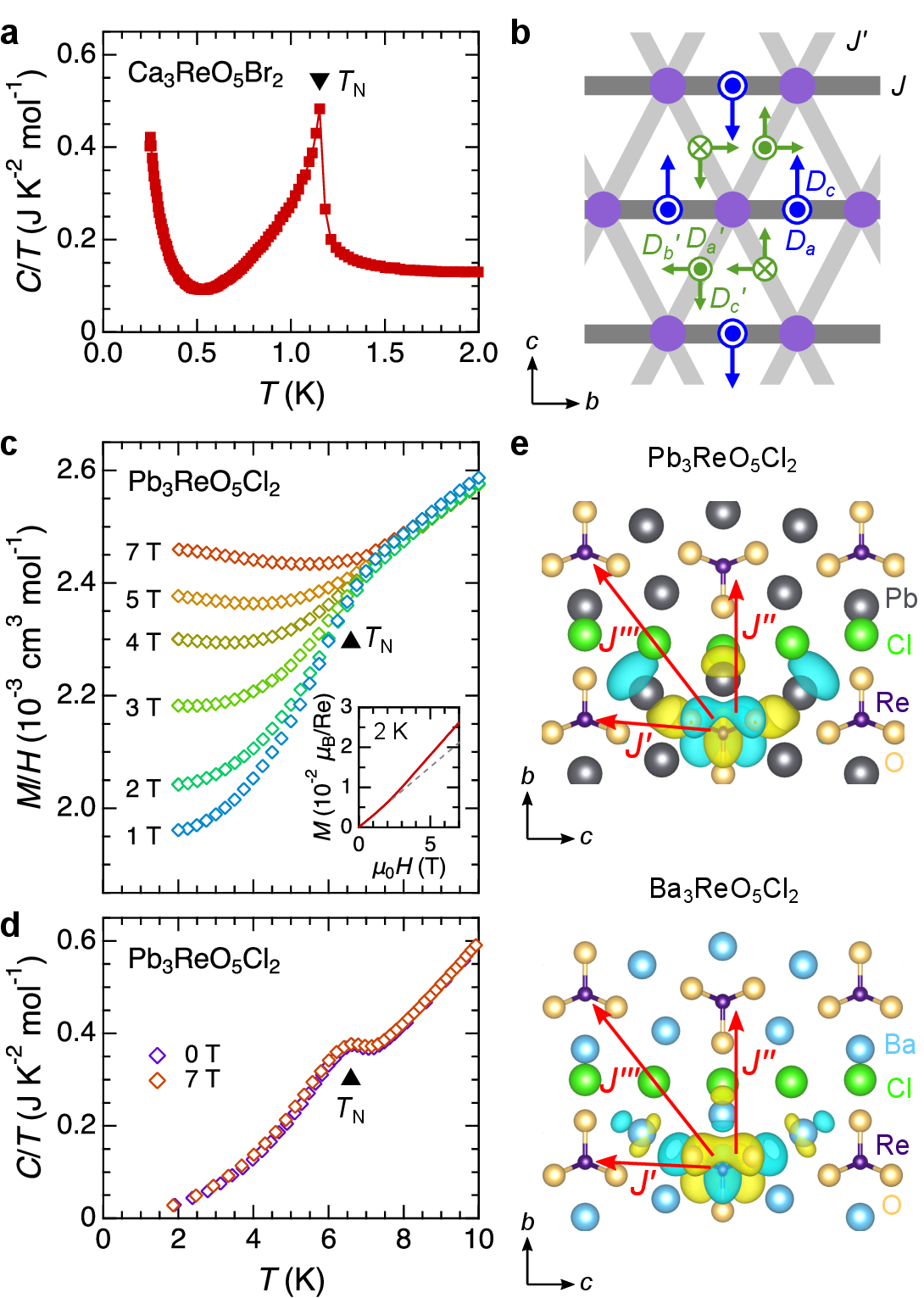}
\caption{{\bf Magnetic transition and its microscopic origin in CROB and PROC.} {\bf a} Temperature dependence of the heat capacity divided by temperature $C/T$ measured at 0~T using $^{3}$He for CROB. {\bf b} Schematic of symmetrically-allowed DM vectors in one ATL layer in CROB \cite{2010_Sta}. $D_{i}$ ($D_{i}'$) ($i = a, b, c$) indicates the $i$ component of the DM vector on the $J$ ($J'$) bond. In the neighboring ATL layer, all the DM components except for $D_{c}$ orient in the opposite direction. {\bf c, d} Temperature dependence of the magnetic susceptibility $M/H$ measured at various magnetic fields ({\bf c}) and the heat capacity divided by temperature $C/T$ at 0 and 7~T ({\bf d}) for PROC. The inset of panel {\bf c} shows a magnetization curve at 2~K measured in a static magnetic field. The gray dashed line is a guide for the eye to clarify the slope change of the magnetization curve. {\bf e} Schematic of the Wannier orbitals around the Pb/Sr and Re atoms for PROC (top) and BROC (bottom) revealed by the first-principles calculations. The paths of intralayer exchange $J'$ and interlayer exchange $J''$ and $J'''$ couplings are depicted by red arrows.}
\label{Fig3}
\end{figure}

\vspace{-0.2cm}
\subsection*{Magnetic phase transition in Ca$_{3}$ReO$_{5}$Br$_{2}$ and Pb$_{3}$ReO$_{5}$Cl$_{2}$}
\vspace{-0.2cm}

In real spin systems, deviations from an ideal spin model more or less exist.
The ground state of the ATL quantum antiferromagnet is highly sensitive to weak perturbations such as the DM interaction and interlayer exchange couplings \cite{2010_Sta}.
For instance, Cs$_{2}$CuCl$_{4}$ undergoes a magnetic transition into a spiral LRO phase with a magnetic propagation vector of $q = (0, 0.475, 0)$ \cite{1996_Col}, which is attributed to a combined effect of magnetic frustration and the DM interaction \cite{2011_Pov}.

As for {\it A}$_{3}$ReO$_{5}${\it X}$_{2}$, the DM interaction is prohibited for {\it A} = Sr, Ba, and Pb with the Type II structure (Supplementary Note 3), whereas this is not the case for {\it A} = Ca with the Type I structure.
Indeed, a neutron scattering experiment on CROC confirmed the appearance of a spiral LRO with $q = (0, 0.465, 0)$ below $T_{\rm N} = 1.13$~K \cite{2019_Hir, 2020_Naw}.
Therefore, a similar magnetic transition is expected for CROB at lower temperatures, though no anomaly is observed in the magnetic susceptibility and heat capacity data above 2~K for CROB as well as SROB and BROB (Fig.~\ref{Fig2}).
Figure~\ref{Fig3}a shows the temperature dependence of $C/T$ measured down to 0.24~K for CROB.
A clear lambda-shaped peak is observed at $T_{\rm N} = 1.15$~K, indicating a second-order transition.
Figure~\ref{Fig3}b illustrates the DM vectors in one ATL layer allowed for the Type I structure (space group $Pnma$) \cite{2010_Sta}.
As discussed in Ref.~\cite{2010_Sta}, the $c$ component of the DM vector on the intrachain bond $D_{c}$ and the $a$ component of the DM vector on the diagonal bond $D_{a}'$ are most relevant in the ground state selection.
For Cs$_{2}$CuCl$_{4}$, it is proposed that the contribution of $D_{a}'$ is significant \cite{2002_Col}, resulting in the spiral moment lying almost within the $bc$ plane \cite{1996_Col}.
On the other hand, a recent electron-spin-resonance (ESR) study suggested that the intrachain DM interaction $D_{c}$ is the primary source for stabilizing the spiral LRO in CROC, as indicated by the observed shift of the soft mode in the disordered phase \cite{2022_Zvy}.
Despite the slightly weaker exchange interactions in CROB than in CROC, they exhibit nearly the same transition temperature, suggesting that the intrachain DM interaction is slightly stronger in CROB.

In addition, we observe a magnetic transition at $T_{\rm N} = 6.6$~K for PROC.
As shown in Fig.~\ref{Fig3}c, the $M/H$--$T$ curve at 0.1~T exhibits a sudden drop below $T_{\rm N}$, indicating an AFM phase transition.
$M/H$ below $T_{\rm N}$ gradually rises with increasing an applied magnetic field up to 7~T, suggesting a spin-flop transition (see also the $M$--$H$ curve at 2~K shown in the inset of Fig.~\ref{Fig3}c).
The $C/T$--$T$ curve exhibits a broad lambda-shaped anomaly indicative of a second-order transition, as shown in Fig.~\ref{Fig3}d.
This anomaly persists at almost the same temperature until 7~T, indicating that the AFM or field-induced spin-canted phases are robust against a higher magnetic field.
The magnetic contribution of $C$ is found to be proportional to $T^{3}$ below $T_{\rm N}$ (Supplementary Note 4), suggesting the presence of 3D magnetic correlations, i.e., interlayer exchange couplings.
This is confirmed by the first-principles calculations: the strengths of interlayer exchange couplings, $J''$ and $J'''$, defined as shown in Fig.~\ref{Fig3}e, are estimated to $J''/J = 0.006$ and $J'''/J = 0.190$ for PROC, whereas both $J''/J$ and $J'''/J$ are much smaller than 0.01 for other compounds (for details, see Supplementary Note~5).
The relatively strong $J'''$ coupling can be attributed to the electron hopping via Pb $6p$ orbital.

\begin{figure}[t]
\centering
\includegraphics[width=\linewidth]{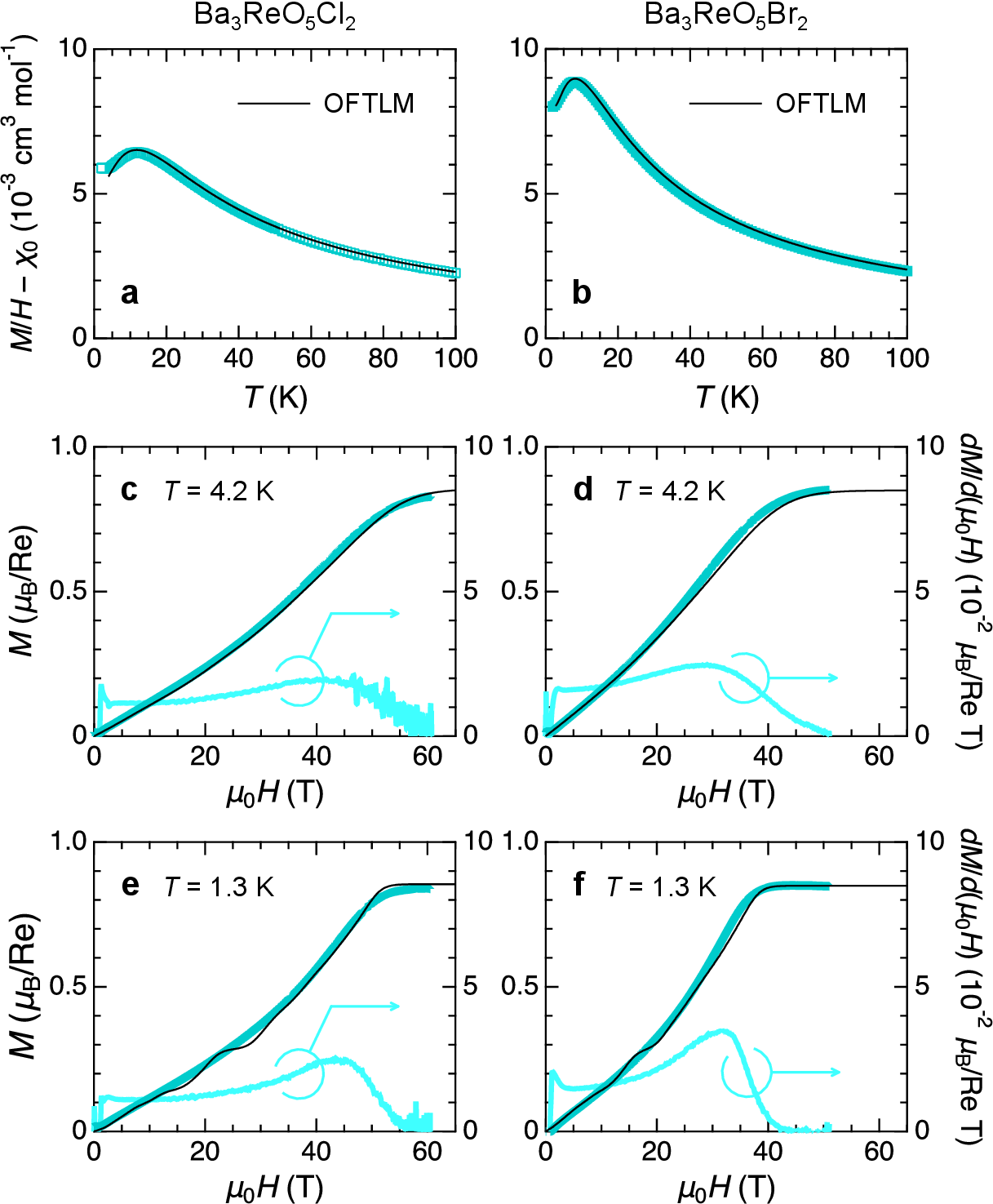}
\caption{{\bf Comparison of magnetization data of Ba$_{3}$ReO$_{5}${\mbox {\boldmath $X$}}$_{2}$ with theoretical calculations based on the OFTLM.} Panels {\bf a} and {\bf b} show the temperature dependence of the magnetic susceptibility measured at 7~T \cite{2020_Hir}, where a $T$-independent component $\chi_{0}$ is subtracted. Panels {\bf c}--{\bf f} show quasi-isothermal magnetization curves in the field-increasing process measured at 4.2~K ({\bf c}, {\bf d}) and 1.3~K ({\bf e}, {\bf f}) using a non-destructive pulsed magnet. Here, the field derivative of the magnetization $dM/d(\mu_{0}H)$ is displayed in the right axis. Experimental curves are shown by cyan lines, and theoretical curves fitted to each experimental data are shown by black lines. Theoretical $M$--$H$ curves represent isothermal curves at temperatures corresponding to the experimental ones. The fitting parameters for each compound are summarized in Table~\ref{Tab2}.}
\label{Fig4}
\end{figure}

\begin{figure*}[t]
\centering
\includegraphics[width=\linewidth]{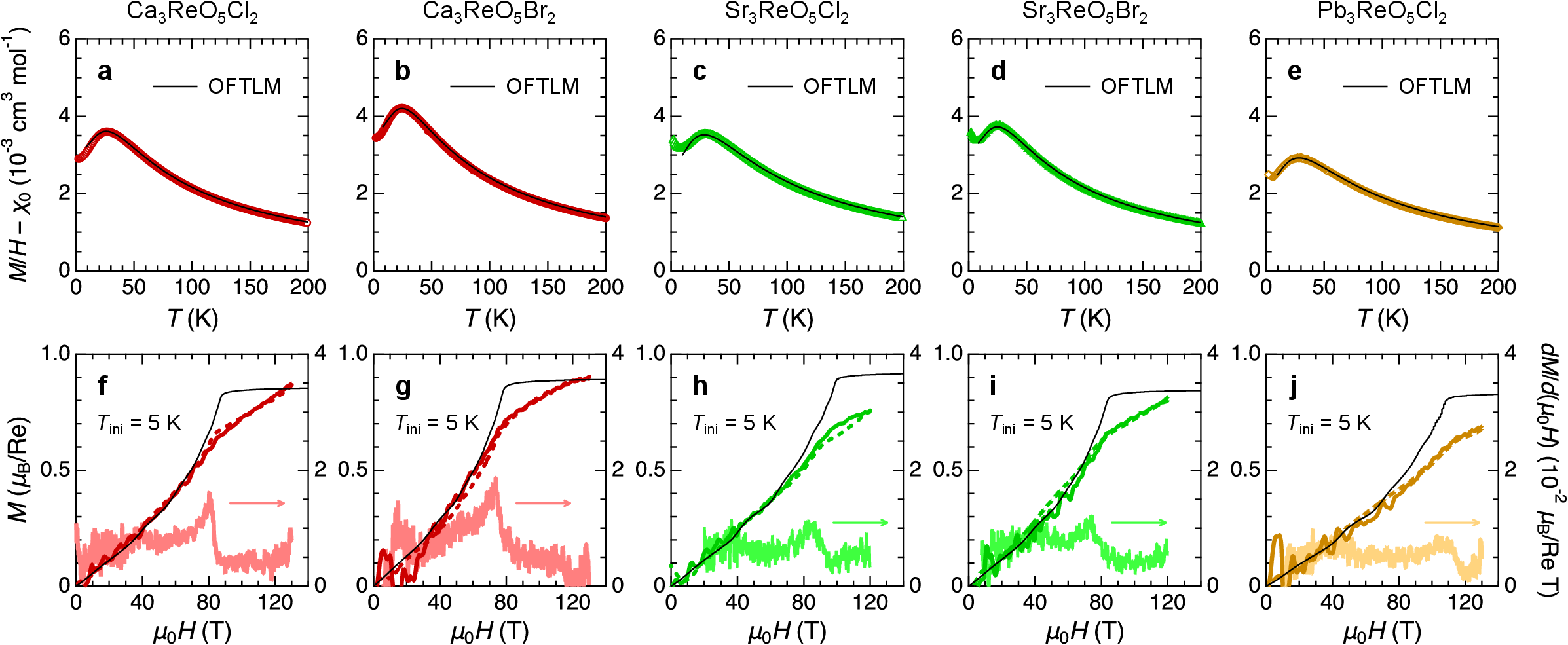}
\caption{{\bf Comparison of magnetization data of {\mbox {\boldmath $A$}}$_{3}$ReO$_{5}${\mbox {\boldmath $X$}}$_{2}$ ({\mbox {\boldmath $A$}} = Ca, Sr, Pb) with theoretical calculations based on the OFTLM.} Panels {\bf a}--{\bf e} show the temperature dependence of the magnetic susceptibility measured at 7~T \cite{2019_Hir, 2020_Hir}, where a $T$-independent component $\chi_{0}$ is subtracted. Panels {\bf f}--{\bf j} show quasi-isentropic magnetization curves in the field-increasing (solid lines) and field-decreasing (dashed lines) processes measured at the initial temperature of $T_{\rm ini} = 5$~K using a single-turn coil megagauss generator. The field derivative of the magnetization $dM/d(\mu_{0}H)$ in the field-decreasing process is displayed in the right axis. Experimental curves are shown by colored lines, and theoretical curves fitted to each experimental data are shown by black lines. Theoretical $M$--$H$ curves represent isentropic curves for $T_{\rm ini} = 5$~K. The fitting parameters for each compound are summarized in Table~\ref{Tab2}.}
\label{Fig5}
\end{figure*}

\vspace{-0.2cm}
\subsection*{High-field magnetization processes}
\vspace{-0.2cm}

Here, we move on to the in-field properties of all the seven {\it A}$_{3}$ReO$_{5}${\it X}$_{2}$ families, which are useful for estimating the strengths of $J$ and $J'$.
Figure~\ref{Fig4} summarizes the magnetization data of BROC and BROB.
The magnetic susceptibility data for BROC (Fig.~\ref{Fig4}a) are taken from Ref.~\cite{2020_Hir}, and those for BROB (Fig.~\ref{Fig4}b) are identical to Fig.~\ref{Fig2}c.
The $M$--$H$ curves in Fig.~\ref{Fig4}c--f are quasi-isothermal processes obtained using a non-destructive pulsed magnet.
The saturation field at 1.3~K is $\mu_{0}H_{\rm sat} = 55(1)$~T for BROC and 42(1)~T for BROB.
The $M$--$H$ curves exhibit a concave shape followed by an inflection point immediately below $H_{\rm sat}$, which is a typical behavior of low-dimensional quantum magnets.

Figure~\ref{Fig5} summarizes the magnetization data of CROC, CROB, SROC, SROB, and PROC from the left panel.
The magnetic susceptibility data for CROC (Fig.~\ref{Fig5}a) and SROC (Fig.~\ref{Fig5}c) are taken from Refs.~\cite{2019_Hir} and \cite{2020_Hir}, respectively.
Those for CROB (Fig.~\ref{Fig5}b), SROB (Fig.~\ref{Fig5}d), and PROC (Fig.~\ref{Fig5}e) are identical to Fig.~\ref{Fig2}a, b, and d, respectively.
The $M$--$H$ curves in Fig.~\ref{Fig5}f--j are quasi-adiabatic processes with the initial temperature of $T_{\rm ini} = 5$~K obtained using a single-turn coil system, i.e., a semi-destructive pulsed magnet.
The absolute values of $M$ are calibrated using the $M$--$H$ curves at 4.2~K obtained in a non-destructive pulsed magnet (Supplementary Note 6).
For all the $M$--$H$ curves, an inflection point indicative of the nearly saturation is observed at $\mu_{0}H_{\rm sat} = 83(3)$~T for CROC, 78(3)~T for CROB, 95(5)~T for SROC, 80(5)~T for SROB, and 115(5)~T for PROC.

\vspace{-0.2cm}
\subsection*{Comparison with theoretical calculations based on the OFTLM}
\vspace{-0.2cm}

The accurate estimation of $J$ and $J'$ in the quantum ATL antiferromagnet is generally challenging.
One might consider achieving this by relating the Weiss temperature and the saturation field to the exchange interactions using the equations $\Theta_{\rm W} = -(J/2 + J')$ and $g\mu_{\rm B}H_{\rm sat} = 2J(1 + J'/2J)^2$, respectively \cite{2005_Zhe, 2014_Zvy}.
However, the experimentally-determined $\Theta_{\rm W}$ tends to underestimate the actual value, especially for small $J'/J$, even when performing the Curie-Weiss fit to the magnetic susceptibility data in a high-temperature range \cite{2005_Zhe}.
Besides, $H_{\rm sat}$ is also highly temperature-dependent.

Recently, the OFTLM was developed \cite{2020_Mor} and applied to the calculation of the magnetic susceptibility and magnetization process in the quantum ATL antiferromagnet \cite{2022_Mor}.
This technique has an advantage over the standard FTLM \cite{1994_Jak} in accurately evaluating the magnetic entropy at low temperatures, which is crucial for the calculation of the adiabatic magnetization process \cite{2022_Mor}.
Here, we adopt the OFTLM to the Hamiltonian of the quantum ATL antiferromagnet in a magnetic field,
\begin{equation}
\label{Eq1}
{\mathcal{H}}=J\sum_{\langle i, j\rangle} {\mathbf S}_{i} \cdot {\mathbf S}_{j}+J'\sum_{\langle i, k\rangle} {\mathbf S}_{i} \cdot {\mathbf S}_{k}-h\sum_{i}S_{i}^{z},
\end{equation}
where ${\mathbf S}_{i}$ denotes a spin-1/2 operator localized on the ATL, ${\langle i, j\rangle}$ and ${\langle i, k\rangle}$ indicate pairs of neighboring sites along the intrachain and interchain directions, respectively, and $h$ is a magnetic field applied along the $z$ axis.
Further details of the calculation are found in the Method section.
Importantly, the peak temperature of the magnetic susceptibility $T_{\rm p}$ normalized to $J$ shifts to a higher temperature side as $J'/J$ decreases \cite{2005_Zhe, 2022_Mor}.
Thanks to this feature, the strengths of $J$ and $J'$ can be uniquely determined once we reveal experimental values of the Land\'{e} $g$-factor, $T_{\rm p}$, and $H_{\rm sat}$, which are listed in Table.~\ref{Tab2}.
Here, we set the $g$-value using a relation $\mu_{\rm eff} = g\sqrt{S(S+1)}$ with $S = 1/2$, where $\mu_{\rm eff}$ is estimated from the Curie-Weiss fit (Table~\ref{Tab1}).

For BROC and BROB, all the magnetization data can be simultaneously reproduced by theoretical curves with a single parameter set, as shown in Fig.~\ref{Fig4}: $J/k_{\rm B} = 20$~K and $J'/J = 0.40$ for BROC, and $J/k_{\rm B} = 14$~K and $J'/J = 0.45$ for BROB.
$J'/J$ for BROB is almost the same with that for BROC.
This is reasonable given that the increase rates of $d$ and $d'$ by the substitution of Cl with Br are almost the same.
No signatures of a 1/3-magnetization plateau are observed for both compounds in contrast to the theoretical $M$--$H$ curves at 1.3~K (Note that fluctuations in the theoretical curves on the low-field side below the 1/3-magnetization plateau are caused by finite-size effects).
We infer that the appearance of the 1/3-magnetization plateau is blurred by averaging over the $M$--$H$ curves on randomly-oriented polycrystals in the present experiments using polycrystalline powder samples.
Indeed, a previous ESR experiment on a CROC single crystal revealed a weak anisotropy: $g = 1.88, 1.85$ and 1.92 for $H \parallel a, b$, and $c$, respectively \cite{2020_Naw}.

Similarly, we calculate magnetic susceptibility and adiabatic magnetization curves for the rest five compounds, as shown in Fig.~\ref{Fig5}.
The exchange parameter set which can reproduce the $M/H$--$T$ curve as well as the saturation field $H_{\rm sat}$ is obtained for each compound as follows: $J/k_{\rm B} = 42$~K and $J'/J = 0.25$ for CROC, $J/k_{\rm B} = 39.5$~K and $J'/J = 0.25$ for CROB, $J/k_{\rm B} = 47$~K and $J'/J = 0.35$ for SROC, $J/k_{\rm B} = 40$~K and $J'/J = 0.25$ for SROB, and $J/k_{\rm B} = 46.5$~K and $J'/J = 0.35$ for PROC.
The deviation between the experimental $M$--$H$ curves and theoretical ones in the high-field region would be mainly due to the imperfect background subtraction in the experimental data (Note that the detection sensitivity of the magnetization pickup coil is significantly reduced near the maximum field).
Given that the error in the experimental $H_{\rm sat}$ is approximately $\pm 5$~T, the error in the estimated value of $J'/J$ is at most $\pm 0.05$.

\begin{table}[t]
\renewcommand{\arraystretch}{1.2}
\caption{Experimental values of the peak temperature of the magnetic susceptibility $T_{\rm p}$, the saturation magnetic field $H_{\rm sat}$, and the Land\'{e} $g$ factor as well as the exchange couplings $J$ and $J'$ estimated based on theoretical calculations using the orthogonalized finite-temperature Lanczos method in the seven {\it A}$_{3}$ReO$_{5}${\it X}$_{2}$ compounds.}
\begin{tabular}{cccccccc} \hline\hline
~~ & ~~~CROC~~~ & ~~~SROC~~~ & ~~~BROC~~~ & ~~~PROC~~~ \\ \hline
~~~$T_{\rm p}$~(K)~~~ & 27 & 29 & 12 & 29 \\
~~~$\mu_{0}H_{\rm sat}$~(T)~~~ & 83(3) & 95(5) & 55(1) & 115(5) \\
~~$g$~~ & 1.776 & 1.903 & 1.709 & 1.724 \\
~~$J/k_{\rm B}$~(K)~~ & 42 & 47 & 20 & 46.5 \\
~~$J'/J$~~ & 0.25 & 0.35 & 0.40 & 0.35 \\
\hline\hline
~~ & ~~~CROB~~~ & ~~~SROB~~~ & ~~~BROB~~~ & \\ \hline
~~~$T_{\rm p}$~(K)~~~ & 24 & 25 & 8.0 & \\
~~~$\mu_{0}H_{\rm sat}$~(T)~~~ & 78(3) & 80(5) & 42(1) & \\
~~$g$~~ & 1.858 & 1.760 & 1.699 & \\
~~$J/k_{\rm B}$~(K)~~ & 39.5 & 40 & 14 & \\
~~$J'/J$~~ & 0.25 & 0.25 & 0.45 & \\\hline\hline
\end{tabular}
\label{Tab2}
\end{table}

\section*{Discussion}

We demonstrate that all three previously reported compounds {\it A}$_{3}$ReO$_{5}$Cl$_{2}$ ({\it A} = Ca, Sr, Ba) \cite{2017_Hir, 2020_Hir} can accommodate the substitution of Cl with Br.
The halide substitution results in the elongation of both $d$ and $d'$ in the ATL composed of Re$^{6+}$ ions (Fig.~\ref{Fig1}c).
These elongations lead to a weakening of the exchange couplings $J$ and $J'$, as indicated by the decrease in $|\Theta_{\rm W}|$ as well as the increase in $\alpha$ ($\propto J^{-1}$) (Table~\ref{Tab1}).
Notably, this tendency is in contrast with another model compound of an ATL quantum antiferromagnet Cs$_{2}$CuCl$_{4}$, where the substitution of Cl with Br leads to the enhancement of $J$ and $J'$ \cite{2014_Zvy}.
We also reveal that the anisotropy $J'/J$ in {\it A}$_{3}$ReO$_{5}${\it X}$_{2}$ ranges from 0.25 to 0.45, although the low-temperature magnetism is commonly characterized by one-dimensionalization irrespective of the types of the {\it A} cation and {\it X} anion.
The halide substitution does not significantly modify $J'/J$ for {\it A} = Ca and Ba, while a substantial decrease in $J'/J$ (from 0.35 to 0.25) is found for $A$ = Sr (Table~\ref{Tab2}).
This can be attributed to a relatively larger increase in $d'$ ($\sim$1.7\%) compared to $d$ ($\sim$1.2\%) for $A$ = Sr (Fig.~\ref{Fig1}c).

The most fascinating aspect of {\it A}$_{3}$ReO$_{5}${\it X}$_{2}$ is the controllability of perturbative interactions, which is responsible for selecting the ground state.
CROC and CROB undergo a magnetic transition at $T_{\rm N} = 1.13$ \cite{2019_Hir} and 1.15~K (Fig.~\ref{Fig3}a), respectively, due to the uniform DM interaction allowed by the space group of the crystal lattice ($Pnma$).
The identification of a magnetic structure in CROB, as well as in CROC, is an important issue to be addressed in future works.
Theoretically, the application of a magnetic field on the ATL quantum antiferromagnet with the DM interaction can induce successive phase transitions associated with the emergence of versatile LRO phases, such as commensurate AFM and incommensurate cone phases, aside from the low-field spiral and forced ferromagnetic phases \cite{2007_Sta, 2010_Sta}.
Indeed, Cs$_{2}$CuCl$_{4}$ exhibits multiple phase transitions for $H \parallel b$ and $c$ \cite{2006_Tok}, and interestingly, the phase diagram becomes more complicated under pressure \cite{2019_Zvy}.
However, the strength of the interlayer exchange ($J''/J = 0.045$) is comparable to that of the DM interaction ($D/J = 0.05$) in Cs$_{2}$CuCl$_{4}$ \cite{2002_Col}, making assignments of several in-field phases somehow puzzling.
For CROC, the interlayer exchange couplings ($J''/J \approx 0.0007$) \cite{2019_Hir} is much weaker than the DM interaction ($D/J \approx 0.23$) \cite{2022_Zvy}, so that the role of $J''$ could be safely neglected.
Detailed magnetization measurements on CROC and CROB single crystals below $T_{\rm N}$ will explicitly elucidate the effect of the DM interaction on the field-induced phases of the ATL quantum antiferromagnet.
Furthermore, PROC undergoes a magnetic transition at $T_{\rm N} = 6.6$~K due to the interlayer exchange couplings via the Re--O--Pb--O--Re superexchange path, which arises from the hybridization of the O $2p$ and Pb $6p$ orbitals.
The exchange paths of the interlayer $J''$ and $J'''$ couplings in PROC (Fig.~\ref{Fig3}e) are more complex than those considered in the previous theoretical work \cite{2010_Sta}.
It is necessary to construct a new theoretical model in line with the spin Hamiltonian of PROC.
The rest four {\it A}$_{3}$ReO$_{5}${\it X}$_{2}$ compounds with {\it A} = Sr or Ba are close to the ideal ATL quantum antiferromagnet free from the interlayer exchange and DM interactions.
Since $J'/J$ for these compounds is smaller than 0.6, the ground states are expected to be a TLL-like gapless QSL, in accordance with previous theories \cite{2006_Yun, 2009_Hei, 2016_Gho}.
The diversity in $J'/J$ values (0.25--0.45) offers an excellent opportunity to meticulously investigate the impact of magnetic frustration on the TLL state.
Investigating the magnetic excitation as well as the dynamical magnetic property at extremely low temperatures would also be promising for understanding the essential nature of the ground state of the ATL quantum antiferromagnet.

\section*{Method}

\subsection*{Sample preparation}

Polycrystalline samples of {\it A}$_{3}$ReO$_{5}${\it X}$_{2}$ ({\it A} = Ca, Sr, Ba, Pb; {\mbox {\it X} = Cl, Br) and an isostructural nonmagnetic compound Pb$_{3}$WO$_{5}$Cl$_{2}$ were synthesized by conventional solid-state reaction.
{\it A}O, {\it A}{\it X}$_{2}$, and ReO$_{3}$ (WO$_{3}$) were mixed in a molar ratio of 2:1:1 in an argon-filled glovebox.
The mixture was pressed into a pellet and wrapped in gold foil, and then sealed in an evacuated quartz tube.
All the samples were sintered twice with an intermediate grinding step.
The sintering conditions for each compound are listed in Supplementary Table~S1. 
Single crystals of {\it A}$_{3}$ReO$_{5}$Br$_{2}$ ({\it A} = Ca, Sr, Ba) with a size of approximately 50~$\mu$m were grown by partially melting a stoichiometric mixture in an evacuated quartz tube at 1050, 1030, 900$^{\circ}$C for 20h for {\it A} = Ca, Sr, and Ba, respectively.

\subsection*{X-ray diffraction measurement and structural analysis}

The synthesized polycrystalline samples were characterized by X-ray diffraction (XRD) measurements with Cu $K\alpha$ radiation using a diffractometer (RINT-2500, Rigaku).
All the polycrystalline samples used for physical property measurements were confirmed to be single phase without any noticeable impurities.
The crystal structures of {\it A}$_{3}$ReO$_{5}$Br$_{2}$ ({\it A} = Ca, Sr, Ba) were determined by single-crystal XRD measurements at room temperature using an R-AXIS RAPID IP diffractometer (Rigaku) with monochromated Mo $K\alpha$ radiation.
The structures were solved by direct methods and refined by full-matrix least-squares methods on $|F^{2}|$ by using the SHELXL2013 software.
The structural analysis of Pb$_{3}$ReO$_{5}$Cl$_{2}$ was performed by the Rietveld method with RIETAN-FP program \cite{2007_Izu} on the basis of the Ba$_{3}$ReO$_{5}$Cl$_{2}$ structure \cite{2020_Hir}.
The powder XRD data used for the structure analysis were collected at room temperature using a SmartLab diffractometer (Rigaku) with Cu $K{\alpha1}$ radiation monochromated by a Ge(111)-Johansson-type monochromator.

\subsection*{Physical property measurements}

Magnetization measurements up to 7 T were performed using a commercial magnetometer (MPMS-3, Quantum Design).
Heat capacity measurements were performed in zero field by a thermal relaxation method using a commercial cryostat equipped with a superconducting magnet (PPMS, Quantum Design) down to 2~K and by a quasi-adiabatic heat-pulse method in a $^{3}$He cryostat down to 0.24~K.
Magnetization measurements up to $\sim$60~T and $\sim$130~T were performed by the induction method using a non-destructive pulsed magnet ($\sim$4~ms duration) and a horizontal single-turn coil system ($\sim$8~$\mu$s duration) \cite{2020_Gen}, respectively.
Polycrystalline samples were used in all the magnetization measurements.
{\it A}$_{3}$ReO$_{5}$Br$_{2}$ ({\it A} = Ca, Sr, Ba) single crystals and Pb$_{3}$ReO$_{5}$Cl$_{2}$ polycrystalline samples were used in the heat capacity measurements.

\subsection*{First-principles calculation}

First-principles calculations are performed based on density functional theory (DFT) using the program package Quantum ESPRESSO \cite{2017_Gia}, which employs plane waves and pseudopotentials to describe the Kohn--Sham orbitals and the crystalline potential, respectively.
We set the plane-wave cutoff for a wave function to 120 Ry, suitable for the optimized norm-conserving pseudopotentials \cite{2013_Ham} provided in the SG15 pseudopotential library \cite{2015_Sch}.
The exchange and correlation effect is described with the Perdew-Burke-Ernzerhof functional \cite{1996_Per} based on the generalized gradient correction.
We perform the Brillouin-zone integral on $5 \times 10 \times 5$ ($10 \times 10 \times 5$) {\it k}-point grids for the charge-density calculation of the Type~I (Type~II) structure by using the optimized tetrahedron method \cite{2014_Kaw}.
The fluctuated position of halogen atoms is fixed to the middle of the fluctuation to make the structure suitable for the DFT calculation.
We computed the maximally localized Wannier function (MLWF) \cite{1997_Mar} associated with the 5$d_{xy}$ orbitals of the Re
atoms.
The onsite screened Coulomb integral ($U$) is computed with the constrained random phase approximation (cRPA) \cite{2009_Miy}.
For the calculations of MLWF and cRPA, we use the RESPACK \cite{2021_Nak} program package.
The susceptibility for the screened Coulomb interaction in Type~I (II) materials is computed with 878 (439) empty bands whose upper limit is 35~eV from the Fermi level.
The plane-wave cutoff for the susceptibility is set to 12~eV.

\subsection*{Orthogonalized finite-temperature Lanczos method}

The finite-temperature Lanczos method (FTLM) is useful for the analysis of frustrated quantum spin systems, though \cite{1994_Jak}.
The orthogonalized finite-temperature Lanczos method (OFTLM) is a more accurate method than the standard FTLM, particularly at low temperatures \cite{2022_Mor, 2020_Mor}.
In the OFTLM, we first calculate several low-lying exact eigenvectors $|\Psi_{i,m} \rangle$ with $N_{V}$ levels, where we define the order of the corresponding eigenvalues \{$E_{i,m}$\} as $E_{0,m} \le E_{1,m} \le \cdots \le E_{N_{V-1},m}$.
By using a normalized random initial vector with $S_{\rm tot}^{z} = m$, $|V_{r,m}\rangle$, we calculate the modulated random vector
\begin{eqnarray} 
|V_{r,m}'\rangle  &=& \left[ I - \sum_{i=0}^{N_V-1} | \Psi_{i,m} \rangle \langle \Psi_{i,m} |  \right] | V_{r,m} \rangle,  \label{r'}
\end{eqnarray}
with normalization
\begin{equation}
|V_{r,m}'\rangle  \Rightarrow \frac{ |V_{r,m}'\rangle }{ \sqrt{\langle V_{r,m}' |V_{r,m}'\rangle} }. \label{r'2}
\end{equation}
The partition function of the OFTLM is obtained as follows:
\begin{equation}
\begin{split}
Z(T,h)_{\rm OFTL} 
&= \sum_{m=-M_{\rm sat}}^{M_{\rm sat}} \left[ \frac{N_{\rm st}^{(m)}-N_V}{R}\sum_{r=1}^{R} \sum_{j=0}^{M_{L-1}}  \right. \\
& \left. e^{-\beta \epsilon^{(r)}_{j,m}(h)} |\langle V_{r,m}' | \psi^{r}_{j,m} \rangle|^2  + \sum_{i=0}^{N_{V-1}} e^{-\beta E_{i,m}(h)} \right],
\label{ZOFTL} 
\end{split}
\end{equation}
where $M_{\rm sat}$ is the saturation magnetization, $N_{\rm st}^{(m)}$ is the dimension of the Hilbert subspace with $S_{\rm tot}^{z} = m$, $R$ denotes the number of random samplings of the OFTLM, $M_{L}$ denotes the dimension of the Krylov subspace, $\beta$ is the inverse temperature 1/$k_{\rm B}T$, and $|\psi^r_{j,m}\rangle$ [$\epsilon^{(r)}_{j,m}(h)$] are the eigenvectors (eigenvalues) in the $M_L$-th Krylov subspace with $S_{\rm tot}^{z} = m$.
The energy $E(T,h)_{\rm OFTL}$, magnetization $M(T,h)_{\rm OFTL}$, magnetic susceptibility $\chi(T)_{\rm OFTL}$, and magnetic entropy $S_{\rm mag}(T,h)_{\rm OFTL}$ are obtained as follows:
\begin{equation}
\begin{split}
E(T,h)_{\rm OFTL}  &= \frac{1}{Z(T,h)_{\rm OFTL}}\sum_{m=-M_{\rm sat}}^{M_{\rm sat}} \left[ \frac{N_{\rm st}^{(m)}-N_V}{R}  \right. \\
      &\times \sum_{r=1}^{R} \sum _{j=0}^{M_L-1} \epsilon^{(r)}_{j,m}(h) e^{-\beta \epsilon^{(r)}_{j,m}(h)} |\langle V_{r,m}' | \psi^{r}_{j,m} \rangle|^2  \\
       &+ \left. \sum_{i=0}^{N_V-1} E_{i,m}(h) e^{-\beta E_{i,m}(h)} \right], 
\label{EOFTL} 
\end{split}
\end{equation}
\begin{equation}
\begin{split}
M(T,h)_{\rm OFTL}  &= \frac{1}{Z(T,h)_{\rm OFTL}}\sum_{m=-M_{\rm sat}}^{M_{\rm sat}} \left[ \frac{N_{\rm st}^{(m)}-N_V}{R}  \right.  \\
                             &\times \sum _{r=1}^{R} \sum _{j=0}^{M_L-1} m e^{-\beta \epsilon^{(r)}_{j,m}(h)} |\langle V_{r,m}' | \psi^{r}_{j,m} \rangle|^2  \\ 
                             &+ \left.  \sum_{i=0}^{N_V-1} m e^{-\beta E_{i,m}(h)} \right], 
\label{MOFTL} 
\end{split}
\end{equation}
\begin{equation}
\begin{split}
  \chi(T)_{\rm OFTL}  &= \frac{1}{k_{\rm B}TZ(T,0)_{\rm OFTL}}\sum_{m=-M_{\rm sat}}^{M_{\rm sat}} \left[ \frac{N_{\rm st}^{(m)}-N_V}{R} \right.  \\
                             &\times \sum _{r=1}^{R} \sum _{j=0}^{M_L-1} m^2 e^{-\beta \epsilon^{(r)}_{j,m}} |\langle V_{r,m}' | \psi^{r}_{j,m} \rangle|^2  \\
                             &+ \left. \sum_{i=0}^{N_V-1} m^2 e^{-\beta E_{i,m}} \right], 
\label{COFTL} 
\end{split}
\end{equation}
\begin{equation} 
  S_{\rm mag}(T,h)_{\rm OFTL}  = \frac{E(T,h)_{\rm OFTL} }{T} - k_{\rm B}\ln Z(T,h)_{\rm OFTL}.
\label{SOFTL} 
\end{equation}
Since the last terms in Eqs.~(\ref{ZOFTL}), (\ref{EOFTL}), (\ref{MOFTL}), and (\ref{COFTL}) are exact values, they are more accurate than those obtained using the standard FTLM, particularly at low temperatures.

In the adiabatic processes, the entropy remains constant while the temperature changes. 
The temperature under the adiabatic processes can be determined based on the following relationship with $h$:
\begin{equation} 
S_{\rm mag}(T,h)_{\rm OFTL}  -  S_{\rm mag}(T_{\rm ini},0)_{\rm OFTL} =0,
\label{adiT-h} 
\end{equation}
where $S_{\rm mag}(T_{\rm ini},0)_{\rm OFTL}$ corresponds to the initial entropy, and $T_{\rm ini}$ represents the initial temperature.
We solved Eq.~(\ref{adiT-h}) for $T$ using the Newton-Raphson method, and then substituted the obtained $T$ into Eq.~(\ref{MOFTL}) to calculate the magnetization under the adiabatic processes.

We performed OFTLM calculations for a cluster of 36 sites under periodic boundary conditions with $R = 10$, $N_V = 4$, and $M_L = 100$.
Our calculations have revealed that there are almost no finite-size effects in the magnetization for $k_{\rm B}T/J > 0.1$ and in the magnetic susceptibility for $k_{\rm B}T/J > 0.2$~\cite{2022_Mor}.
Therefore, the analysis of the magnetic susceptibility and magnetization curves in this study is sufficiently accurate.
 
\section*{Data availability}
The datasets generated during and/or analyzed during the current study are available from the corresponding author upon reasonable request.
The data from the first-principles calculations are available in the ARIM-mdx data repository \cite{Data, 2022_Suz}.

\section*{Acknowledgements}
We thank the Supercomputer Center, the Institute for Solid State Physics, the University of Tokyo for providing us with the CPU time.
This work was partly supported by the Japan Society for the Promotion of Science (JSPS) KAKENHI Grants-In-Aid for Scientific Research (No.~19H05821, No.~20H01858, No.~20J10988, No.~22H01167, No.~23H04860).
M.G. was a postdoctoral research fellow of the JSPS.

\section*{Author Contributions}
M.G. and D.H. conceived and organized the project.
S.K. and D.H. synthesized samples under supervision of Z.H.;
T.Y. performed the XRD measurement and structural analysis.
D.H. and K.D. performed the magnetic susceptibility and heat capacity measurements.
A.M. performed the magnetization measurement in the non-destructive pulsed magnet under supervision of M.G. and K.K.;
M.G. and N.M. performed the magnetization measurement in the single-turn coil system under supervision of Y.K.;
K.M. performed the OFTLM calculations.
M.K. performed the first-principles calculations.
M.G. wrote the manuscript with input comments from all co-authors.

\section*{Competing Interests}
The authors declare no competing financial or non-financial interests.

\section*{Additional Information}
Supplementary Information accompanies this article.


\end{document}


\title{Supplementary Information for\\``Rhenium oxyhalides:
a showcase for anisotropic-triangular lattice quantum antiferromagnets''}

\author{M. Gen}
\email{masaki.gen@riken.jp}
\affiliation{Institute for Solid State Physics, The University of Tokyo, Kashiwa, Chiba 277-8581, Japan}
\affiliation{RIKEN Center for Emergent Matter Science (CEMS), Wako 351-0198, Japan}

\author{D. Hirai}
\affiliation{Institute for Solid State Physics, The University of Tokyo, Kashiwa, Chiba 277-8581, Japan}
\affiliation{Department of Applied Physics, Nagoya University, Nagoya 464-8603, Japan}

\author{K. Morita} 
\affiliation{Department of Physics, Faculty of Science and Technology, Tokyo University of Science, Chiba 278-8510, Japan}

\author{S. Kogane}
\affiliation{Institute for Solid State Physics, The University of Tokyo, Kashiwa, Chiba 277-8581, Japan}

\author{N. Matsuyama}
\affiliation{Institute for Solid State Physics, The University of Tokyo, Kashiwa, Chiba 277-8581, Japan}

\author{T. Yajima}
\affiliation{Institute for Solid State Physics, The University of Tokyo, Kashiwa, Chiba 277-8581, Japan}

\author{M. Kawamura}
\affiliation{Institute for Solid State Physics, The University of Tokyo, Kashiwa, Chiba 277-8581, Japan}
\affiliation{Information Technology Center, The University of Tokyo, Tokyo 113-8658, Japan}

\author{K. Deguchi}
\affiliation{Department of Physics, Nagoya University, Nagoya 464-8602, Japan}

\author{A. Matsuo}
\affiliation{Institute for Solid State Physics, The University of Tokyo, Kashiwa, Chiba 277-8581, Japan}

\author{K. Kindo}
\affiliation{Institute for Solid State Physics, The University of Tokyo, Kashiwa, Chiba 277-8581, Japan}

\author{Y. Kohama}
\affiliation{Institute for Solid State Physics, The University of Tokyo, Kashiwa, Chiba 277-8581, Japan}

\author{Z. Hiroi}
\affiliation{Institute for Solid State Physics, The University of Tokyo, Kashiwa, Chiba 277-8581, Japan}

\maketitle

\onecolumngrid

\vspace{+0.5cm}
This Supplementary Information includes contents as listed below:\\
\\
Note~1.~Synthesis conditions for polycrystalline samples\\
Note~2.~Structural analysis for four novel compounds\\
Note~3.~Absence of Dzyaloshinskii-Moriya interaction for the Type-II structure\\
Note~4.~Magnetic heat capacity below $T_{\rm N}$ for Pb$_{3}$ReO$_{5}$Cl$_{2}$\\
Note~5.~Details of first-principles calculations results\\
Note~6.~Magnetization curves measured using a nondestructive pulsed magnet\\

\vspace{+3.0cm}
\subsection*{\label{Sec1} \large Note 1. Synthesis conditions for polycrystalline samples}

Table~\ref{TabS1} summarizes the synthesis conditions for polycrystalline samples of {\it A}$_{3}$ReO$_{5}${\it X}$_{2}$ ({\it A} = Ca, Sr, Ba, Pb; {\it X} = Cl, Br) and Pb$_{3}$WO$_{5}$Cl$_{2}$ in the solid state reactions.
The heating and cooling rates were set at 200$^{\circ}$C/h for all the syntheses.

\begin{table}[h]
\renewcommand{\thetable}{S\arabic{table}}
\caption{Synthesis conditions for the polycrystalline samples used in this work.}
\vspace{+0.1cm}
\begin{tabular}{c|c|c} \hline\hline
~~~Compound~~~ & ~~~Initial sintering condition~~~ & ~~~Second sintering condition~~~ \\ \hline
Ca$_{3}$ReO$_{5}$Cl$_{2}$ & 800$^{\circ}$C for 24h & 900$^{\circ}$C for 48h \\
Ca$_{3}$ReO$_{5}$Br$_{2}$ & 700$^{\circ}$C for 24h & 750$^{\circ}$C for 24h \\
~~~Sr$_{3}$ReO$_{5}$Cl$_{2}$, Sr$_{3}$ReO$_{5}$Br$_{2}$~~~ & 750$^{\circ}$C for 24h & 800$^{\circ}$C for 48h \\
~~~Ba$_{3}$ReO$_{5}$Cl$_{2}$, Ba$_{3}$ReO$_{5}$Br$_{2}$~~~ & 750$^{\circ}$C for 24h & 800$^{\circ}$C for 48h \\
Pb$_{3}$ReO$_{5}$Cl$_{2}$, Pb$_{3}$WO$_{5}$Cl$_{2}$ & 550$^{\circ}$C for 24h & 570$^{\circ}$C for 24h \\
\hline\hline
\end{tabular}
\label{TabS1}
\end{table}

\newpage
\subsection*{\label{Sec2} \large Note 2. Structural analysis for four novel compounds}

Table~\ref{TabS2} summarizes the crystallographic parameters of the four novel compounds, {\it A}$_{3}$ReO$_{5}$Br$_{2}$ ({\it A} = Ca, Sr, Ba) and Pb$_{3}$ReO$_{5}$Cl$_{2}$.
The structural analysis was performed using the single-crystal XRD data for the three bromides, and the Rietveld analysis was performed using the powder XRD data for Pb$_{3}$ReO$_{5}$Cl$_{2}$.

\vspace{+0.5cm}
\begin{table}[h]
\renewcommand{\thetable}{S\arabic{table}}
\caption{Crystallographic parameters for Ca$_{3}$ReO$_{5}$Br$_{2}$, Sr$_{3}$ReO$_{5}$Br$_{2}$, Ba$_{3}$ReO$_{5}$Br$_{2}$, and Pb$_{3}$ReO$_{5}$Cl$_{2}$.}
\vspace{+0.1cm}
\begin{tabular}{ccccccc} \hline\hline
~~ & ~~ & ~$x$~ & ~$y$~ & ~$z$~ & Occupancy & ~100$U_{\rm iso}$ (\AA$^{2}$)~ \\ \hline
\multicolumn{7}{l}{~Ca$_{3}$ReO$_{5}$Br$_{2}$ ($R = 4.58~\%$, $wR = 10.24~\%$)} \\
\multicolumn{7}{l}{~Space group $Pnma$ (No.~62), $a = 12.1219(14)~\AA$, $b = 5.6447(7)~\AA$, $c = 11.2205(10)~\AA$} \\
~~~~~~~~Ca1~~~~ & ~~~~$4c$~~~~ & ~~~~0.35385(15)~~~~ & ~~~~1/4~~~~ & ~~~~0.84889(14)~~~~ & ~~~~1~~~~ & ~~~~1.14(4)~~~~ \\
~~~~~~~~Ca2~~~~ & ~~~~$4c$~~~~ & ~~~~0.33714(17)~~~~ & ~~~~3/4~~~~ & ~~~~0.59350(11)~~~~ & ~~~~1~~~~ & ~~~~1.04(4)~~~~ \\
~~~~~~~~Ca3~~~~ & ~~~~$4c$~~~~ & ~~~~0.33380(15)~~~~ & ~~~~1/4~~~~ & ~~~~0.34145(16)~~~~ & ~~~~1~~~~ & ~~~~1.09(4)~~~~ \\
~~~~~~~~Re~~~~ & ~~~~$4c$~~~~ & ~~~~0.19045(3)~~~~ & ~~~~1/4~~~~ & ~~~~0.59694(2)~~~~ & ~~~~1~~~~ & ~~~~0.78(2)~~~~ \\
~~~~~~~~O1~~~~ & ~~~~$8d$~~~~ & ~~~~0.2431(4)~~~~ & ~~~~0.0269(6)~~~~ & ~~~~0.7113(3)~~~~ & ~~~~1~~~~ & ~~~~1.07(8)~~~~ \\
~~~~~~~~O2~~~~ & ~~~~$8d$~~~~ & ~~~~0.2392(3)~~~~ & ~~~~0.4784(7)~~~~ & ~~~~0.4817(3)~~~~ & ~~~~1~~~~ & ~~~~1.08(8)~~~~ \\
~~~~~~~~O3~~~~ & ~~~~$4c$~~~~ & ~~~~0.0495(6)~~~~ & ~~~~1/4~~~~ & ~~~~0.5984(4)~~~~ & ~~~~1~~~~ & ~~~~1.33(13)~~~~ \\
~~~~~~~~Br1~~~~ & ~~~~$4c$~~~~ & ~~~~0.52924(7)~~~~ & ~~~~1/4~~~~ & ~~~~0.19156(6)~~~~ & ~~~~1~~~~ & ~~~~1.39(3)~~~~ \\
~~~~~~~~Br2~~~~ & ~~~~$4c$~~~~ & ~~~~0.50835(11)~~~~ & ~~~~3/4~~~~ & ~~~~0.40308(7)~~~~ & ~~~~1~~~~ & ~~~~2.29(3)~~~~ \\ \hline
\multicolumn{7}{l}{~Sr$_{3}$ReO$_{5}$Br$_{2}$ ($R = 2.77~\%$, $wR = 5.13~\%$)} \\
\multicolumn{7}{l}{~Space group $Cmcm$ (No.~63), $a = 5.7145(3)~\AA$, $b = 13.3445(7)~\AA$, $c = 11.3044(6)~\AA$} \\
~~~~~~~~Sr1~~~~ & ~~~~$4c$~~~~ & ~~~~0~~~~ & ~~~~0.38997(8)~~~~ & ~~~~1/4~~~~ & ~~~~1~~~~ & ~~~~1.29(2)~~~~ \\
~~~~~~~~Sr2~~~~ & ~~~~$8f$~~~~ & ~~~~0~~~~ & ~~~~0.15981(6)~~~~ & ~~~~0.46553(6)~~~~ & ~~~~1~~~~ & ~~~~1.563(19)~~~~ \\
~~~~~~~~Re~~~~ & ~~~~$4c$~~~~ & ~~~~1/2~~~~ & ~~~~0.23305(3)~~~~ & ~~~~1/4~~~~ & ~~~~1~~~~ & ~~~~1.088(14)~~~~ \\
~~~~~~~~O1~~~~ & ~~~~$16h$~~~~ & ~~~~0.7273(6)~~~~ & ~~~~0.2745(3)~~~~ & ~~~~0.3639(3)~~~~ & ~~~~1~~~~ & ~~~~1.47(8)~~~~ \\
~~~~~~~~O2~~~~ & ~~~~$4c$~~~~ & ~~~~1/2~~~~ & ~~~~0.1059(6)~~~~ & ~~~~1/4~~~~ & ~~~~1~~~~ & ~~~~2.44(19)~~~~ \\
~~~~~~~~Br1~~~~ & ~~~~$8f$~~~~ & ~~~~1/2~~~~ & ~~~~0.49837(11)~~~~ & ~~~~0.27498(15)~~~~ & ~~~~0.5~~~~ & ~~~~2.29(6)~~~~ \\
~~~~~~~~Br2~~~~ & ~~~~$8e$~~~~ & ~~~~0.0573(3)~~~~ & ~~~~1/2~~~~ & ~~~~1/2~~~~ & ~~~~0.5~~~~ & ~~~~2.98(6)~~~~ \\ \hline
\multicolumn{7}{l}{~Ba$_{3}$ReO$_{5}$Br$_{2}$ ($R = 3.38~\%$, $wR = 4.82~\%$)} \\
\multicolumn{7}{l}{~Space group $Cmcm$ (No.~63), $a = 5.87477(4)~\AA$, $b = 14.19916(11)~\AA$, $c = 11.60519(9)~\AA$} \\
~~~~~~~~Ba1~~~~ & ~~~~$8f$~~~~ & ~~~~0~~~~ & ~~~~0.15430(5)~~~~ & ~~~~0.46319(9)~~~~ & ~~~~1~~~~ & ~~~~0.058(12)~~~~ \\
~~~~~~~~Ba2~~~~ & ~~~~$4c$~~~~ & ~~~~1/2~~~~ & ~~~~0.10984(8)~~~~ & ~~~~3/4~~~~ & ~~~~1~~~~ & ~~~~0.058(12)~~~~ \\
~~~~~~~~Re~~~~ & ~~~~$4c$~~~~ & ~~~~0~~~~ & ~~~~0.26585(6)~~~~ & ~~~~3/4~~~~ & ~~~~1~~~~ & ~~~~0.1~~~~ \\
~~~~~~~~O1~~~~ & ~~~~$16h$~~~~ & ~~~~0.2366(10)~~~~ & ~~~~0.2314(4)~~~~ & ~~~~0.6396(5)~~~~ & ~~~~1~~~~ & ~~~~2.2(2)~~~~ \\
~~~~~~~~O2~~~~ & ~~~~$4c$~~~~ & ~~~~0~~~~ & ~~~~0.3914(8)~~~~ & ~~~~3/4~~~~ & ~~~~1~~~~ & ~~~~2.2(2)~~~~ \\
~~~~~~~~Br1~~~~ & ~~~~$8f$~~~~ & ~~~~0~~~~ & ~~~~0.00700(14)~~~~ & ~~~~0.7624(7)~~~~ & ~~~~0.5~~~~ & ~~~~0.54(8)~~~~ \\
~~~~~~~~Br2~~~~ & ~~~~$8e$~~~~ & ~~~~0.5220(17)~~~~ & ~~~~0~~~~ & ~~~~1/2~~~~ & ~~~~0.5~~~~ & ~~~~1.81(11)~~~~ \\ \hline
\multicolumn{7}{l}{~Pb$_{3}$ReO$_{5}$Cl$_{2}$ ($R = 3.77~\%$, $wR = 3.85~\%$)} \\
\multicolumn{7}{l}{~Space group $Cmcm$ (No.~63), $a = 5.59462(4)~\AA$, $b = 13.26905(8)~\AA$, $c = 10.97535(7)~\AA$} \\
~~~~~~~~Pb1~~~~ & ~~~~$8f$~~~~ & ~~~~0~~~~ & ~~~~0.35012(3)~~~~ & ~~~~0.52789(4)~~~~ & ~~~~1~~~~ & ~~~~1.12(10)~~~~ \\
~~~~~~~~Pb2~~~~ & ~~~~$4c$~~~~ & ~~~~0~~~~ & ~~~~0.89107(4)~~~~ & ~~~~1/4~~~~ & ~~~~1~~~~ & ~~~~1.125(15)~~~~ \\
~~~~~~~~Re~~~~ & ~~~~$4c$~~~~ & ~~~~0~~~~ & ~~~~0.22927(5)~~~~ & ~~~~1/4~~~~ & ~~~~1~~~~ & ~~~~0.425(14)~~~~ \\
~~~~~~~~O1~~~~ & ~~~~$16h$~~~~ & ~~~~0.2346(6)~~~~ & ~~~~0.2682(3)~~~~ & ~~~~0.1300(4)~~~~ & ~~~~1~~~~ & ~~~~1.17(11)~~~~ \\
~~~~~~~~O2~~~~ & ~~~~$4c$~~~~ & ~~~~0~~~~ & ~~~~0.1074(6)~~~~ & ~~~~1/4~~~~ & ~~~~1~~~~ & ~~~~2.1(3)~~~~ \\
~~~~~~~~Cl1~~~~ & ~~~~$4a$~~~~ & ~~~~0~~~~ & ~~~~0~~~~ & ~~~~0~~~~ & ~~~~1~~~~ & ~~~~4.81(14)~~~~ \\
~~~~~~~~Cl2~~~~ & ~~~~$4c$~~~~ & ~~~~0~~~~ & ~~~~0.4752(3)~~~~ & ~~~~1/4~~~~ & ~~~~1~~~~ & ~~~~3.38(11)~~~~ \\ \hline\hline
\end{tabular}
\vspace{+1.0cm}
\label{TabS2}
\end{table}

\subsection*{\label{Sec3} \large Note 3. Absence of Dzyaloshinskii-Moriya interaction for the Type-II structure}

The DM interaction is absent for {\it A} = Sr, Ba, and Pb with the Type~II structure due to the following reasons:\\
(1) An inversion symmetry exists at the center of the Re--Re diagonal $J'$ bonds.\\
(2) There are mirror planes perpendicular to the $a$ and $c$ axes.\\
(3) Two-fold rotation axis along the $b$ axis exists at the center of the Re--Re intrachain $J$ bonds.

\newpage
\subsection*{\label{Sec4} \large Note 4. Magnetic heat capacity below {\mbox {\boldmath $T$}}$_{\rm N}$ for Pb$_{3}$ReO$_{5}$Cl$_{2}$}

\begin{figure}[h]
\centering
\includegraphics[width=0.85\linewidth]{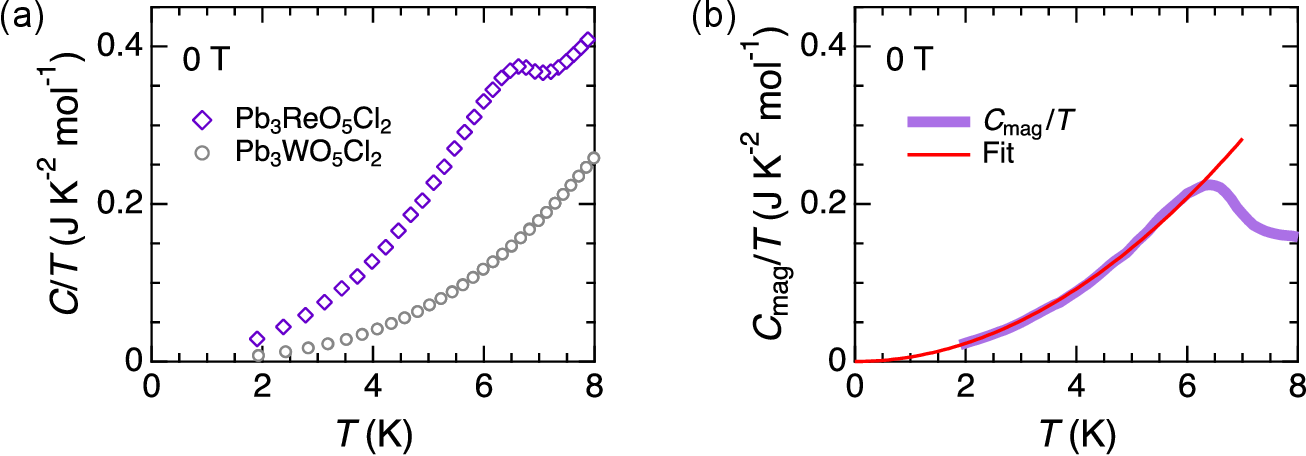}
\renewcommand{\thefigure}{S\arabic{figure}}
\caption{(a) Heat capacity divided by temperature $C/T$ at 0 T for Pb$_{3}$ReO$_{5}$Cl$_{2}$ (diamonds) and Pb$_{3}$WO$_{5}$Cl$_{2}$ (circles). (b) Estimated magnetic heat capacity divided by temperature $C_{\rm mag}/T$ for Pb$_{3}$ReO$_{5}$Cl$_{2}$. The red line represents a fit of $C_{\rm mag}=kT^{3}$ to the experimental data between 2 and 5.5~K.}
\label{FigS1}
\end{figure}

To estimate the lattice heat capacity of Pb$_{3}$ReO$_{5}$Cl$_{2}$, we measured the heat capacity of isostructural nonmagnetic compound Pb$_{3}$WO$_{5}$Cl$_{2}$ in zero field.
Figure~\ref{FigS1}(a) shows $C/T$--$T$ curves of Pb$_{3}$ReO$_{5}$Cl$_{2}$ and Pb$_{3}$WO$_{5}$Cl$_{2}$ at low temperatures.
By subtracting the experimental $C/T$ values of  Pb$_{3}$WO$_{5}$Cl$_{2}$ from those of Pb$_{3}$ReO$_{5}$Cl$_{2}$, the temperature dependence of the magnetic heat capacity, $C_{\rm mag}$, for Pb$_{3}$ReO$_{5}$Cl$_{2}$ was obtained, as shown in Fig.~\ref{FigS1}(b).
Below $T_{\rm N}$, $C_{\rm mag}$ is proportional to $T^{3}$, as indicated by a red fitting curve.
This suggests the presence of 3D magnetic correlations in the magnetically ordered phase of Pb$_{3}$ReO$_{5}$Cl$_{2}$.

\vspace{+0.5cm}

\vspace{+1.0cm}
\subsection*{\label{Sec5} \large Note 5. Details of first-principles calculations results}

In order to evaluate the strengths of the interlayer exchange couplings in {\it A}$_{3}$ReO$_{5}${\it X}, we performed the first-principles calculations based on density functional theory.
Here, we used the structural parameters shown in Table.~\ref{TabS2}.
Note that, for Sr$_{3}$ReO$_{5}$Br$_{2}$ and Ba$_{3}$ReO$_{5}$Br$_{2}$, we fixed the atomic position of Br in the average position of two different Br sites with half occupancy.
Table~\ref{TabS3} summarizes the calculation results of the hopping integrals between two Re$^{6+}$ ions along the intrachain and interchain bonds, $t$ and $t'$, respectively, as well as those between the interlayer bonds, $t''$, and $t'''$, defined for the Type-II structure as shown in Fig.~3e in the main text.
The results for {\it A}$_{3}$ReO$_{5}$Cl$_{2}$ ({\it A} = Ca, Sr, Ba) were already published in Refs.~\cite{2019_Hir, 2020_Hir}.
Considering that the magnetic exchange energy can be related as $J=4t^{2}/U$, where $U$ is the Coulomb integral, it is suggested that the strengths of the interlayer exchange couplings $J''$ and $J'''$ in PROC are one to two orders of magnitude stronger than in other compounds.

\vspace{+0.5cm}
\begin{table}[h]
\renewcommand{\thetable}{S\arabic{table}}
\caption{Hopping integrals in {\it A}$_{3}$ReO$_{5}${\it X} obtained by the first-principles calculations.}
\vspace{+0.1cm}
\begin{tabular}{c|ccccc} \hline\hline
~~~Compound~~~ & ~~~$t$~(eV)~~~ & ~~~$t'$~(eV)~~~ & ~~~$t''$~(eV)~~~ & ~~~$t'''$~(eV)~~~ & ~~~$U$~(eV)~~~ \\ \hline
~~Ca$_{3}$ReO$_{5}$Cl$_{2}$~~ & ~~$-0.050765$~~ & ~~$-0.027532$~~ & N/A & N/A & ~~$3.391358$~~ \\
~~Ca$_{3}$ReO$_{5}$Br$_{2}$~~ & ~~$-0.048177$~~ & ~~$-0.026705$~~ & N/A & N/A & ~~$3.277585$~~ \\
~~Sr$_{3}$ReO$_{5}$Cl$_{2}$~~ & ~~$-0.058905$~~ & ~~$-0.020274$~~ & ~~$0.000002$~~ & ~~$-0.002359$~~ & ~~$3.373051$~~ \\
~~Sr$_{3}$ReO$_{5}$Br$_{2}$~~ & ~~$-0.055679$~~ & ~~$-0.019031$~~ & ~~$-0.000070$~~ & ~~$-0.003175$~~ & ~~$3.302044$~~ \\
~~Ba$_{3}$ReO$_{5}$Cl$_{2}$~~ & ~~$-0.040984$~~ & ~~$-0.022597$~~ & ~~$-0.000587$~~ & ~~$-0.002715$~~ & ~~$3.159830$~~ \\
~~Ba$_{3}$ReO$_{5}$Br$_{2}$~~ & ~~$-0.051275$~~ & ~~$-0.020229$~~ & ~~$0.001340$~~ & ~~$-0.003811$~~ & ~~$2.912788$~~ \\
~~Pb$_{3}$ReO$_{5}$Cl$_{2}$~~ & ~~$-0.064862$~~ & ~~$-0.020121$~~ & ~~$0.005146$~~ & ~~$-0.028263$~~ & ~~$2.112172$~~ \\
\hline\hline
\end{tabular}
\vspace{+1.0cm}
\label{TabS3}
\end{table}

\newpage
\subsection*{\label{Sec6} \large Note 6. Magnetization curves measured using a nondestructive pulsed magnet}

\begin{figure}[h]
\centering
\includegraphics[width=0.5\linewidth]{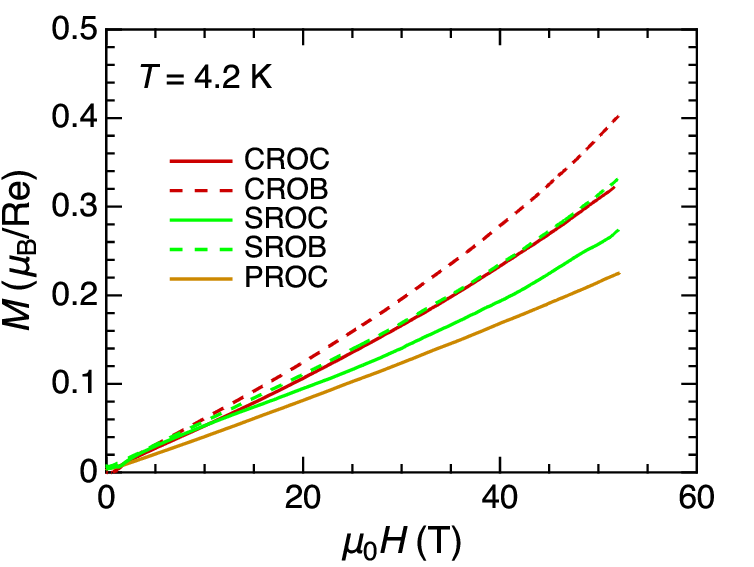}
\renewcommand{\thefigure}{S\arabic{figure}}
\caption{Quasi-isothermal magnetization curves for five {\it A}$_{3}$ReO$_{5}${\it X}$_{2}$ compounds measured at 4.2~K using a non-destructive pulsed magnet.}
\label{FigS2}
\end{figure}

Figure~\ref{FigS2} shows the $M$--$H$ curves of five compounds, {\it A}$_{3}$ReO$_{5}${\it X}$_{2}$ ({\it A} = Ca, Sr, Pb; {\it X} = Cl, Br), measured at 4.2~K using a non-destructive pulsed magnet.
These data are used for calibrating the absolute value of $M$ for the $M$--$H$ curves measured using a single-turn coil system (shown in Fig.~5f--j in the main text).

\vspace{+0.5cm}
